\documentclass{emulateapj}
\usepackage{ltgtsim,epsfig}

\journalinfo{Accepted for publication in \apj, May 9, 2005}
\submitted{Accepted for publication in \apj, May 9, 2005}

\newcommand{\ls}{\lambda_{\rm s}}
\newcommand{\lsthree}{\lambda_{\rm s3}}
\newcommand{\lj}{\lambda_{\rm J}}
\newcommand{\ljzero}{\lambda_{\rm J0}}
\newcommand{\cs}{c_{\rm s}}

\newcommand{\tff}{t_{\rm ff}}
\newcommand{\calt}{\mathcal{T}}
\newcommand{\calm}{\mathcal{M}}
\newcommand{\calw}{\mathcal{W}}

\newcommand{\sr}{\sigma_{\rho}}
\newcommand{\ec}{\epsilon_{\rm core}}
\newcommand{\xc}{x_{\rm crit}}
\newcommand{\avir}{\alpha_{\rm vir}}
\newcommand{\sfrff}{\mbox{SFR}_{\rm ff}}
\newcommand{\sfr}{\dot{M}_*}
\newcommand{\cfrff}{\mbox{CFR}_{\rm ff}}
\newcommand{\msun}{\mbox{M}_{\odot}}

\newcommand{\mphi}{M_{\Phi}}

\newcommand{\tcr}{t_{\rm cr}}

\newcommand{\phix}{\phi_{x}}
\newcommand{\phit}{\phi_{t}}
\newcommand{\phip}{\phi_{P}}
\newcommand{\phimp}{\phi_{\rm mp}}
\newcommand{\Ssfr}{\dot{\Sigma}_{*}}
\newcommand{\Smol}{\Sigma_{\rm mol}}
\newcommand{\Stot}{\Sigma_{\rm tot}}
\newcommand{\Sg}{\Sigma_{\rm g}}
\newcommand{\sg}{\sigma_{\rm g}}
\newcommand{\Scl}{\Sigma_{\rm cl}}
\newcommand{\phipbar}{\phi_{\overline{P}}}
\newcommand{\vrot}{v_{\rm rot}}
\newcommand{\hg}{h_{\rm g}}
\newcommand{\rhog}{\rho_{\rm g}}
\newcommand{\mcl}{M_{\rm cl}}
\newcommand{\rcl}{R_{\rm cl}}
\newcommand{\scl}{\sigma_{\rm cl}}
\newcommand{\rhocl}{\rho_{\rm cl}}
\newcommand{\pcl}{P_{\rm cl}}
\newcommand{\fg}{f_{\rm g}}

\newcommand{\Sgtwo}{\Sigma_{\rm g,2}}

\newcommand{\tdyn}{\tau_{\rm dyn}}

\newcommand{\Omzero}{\Omega_{0}}
\newcommand{\Qonefive}{Q_{1.5}}
\newcommand{\fGMC}{f_{\rm GMC}}

\newcommand{\tffg}{t_{\rm ff-g}}
\newcommand{\phirho}{\phi_{\rho}}
\newcommand{\phipbarsix}{\phi_{\overline{P},6}}

\shorttitle{Turbulence-Regulated Star Formation}
\shortauthors{Krumholz \& McKee}

\begin{document}


\title{A General Theory of Turbulence-Regulated Star Formation, From
Spirals to ULIRGs}


\author{Mark R. Krumholz}
\affil{Physics Department, University of California, Berkeley,
Berkeley, CA 94720}
\email{krumholz@astron.berkeley.edu}

\author{Christopher F. McKee}
\affil{Departments of Physics and Astronomy, University of California,
Berkeley, Berkeley, CA 94720}
\email{cmckee@astron.berkeley.edu}


\begin{abstract}
We derive an analytic prediction for the star formation rate in
environments ranging from normal galactic disks to
starbursts and ULIRGs in terms of the observables of those
systems. Our calculation is based on three premises: (1) star
formation occurs in virialized molecular clouds that are
supersonically turbulent; (2) the density distribution within
these clouds is lognormal, as expected for supersonic isothermal
turbulence; (3) stars form in any sub-region of a cloud that is so
overdense that its gravitational potential energy exceeds the energy
in turbulent motions. We show that a theory based on this model is
consistent with simulations and with the observed star formation rate
in the Milky Way. We use our theory to derive the Kennicutt-Schmidt
Law from first principles, and make other predictions that can be
tested by future observations. We also provide an algorithm for
estimating the star formation rate that is suitable for inclusion in
numerical simulations.
\end{abstract}


\keywords{galaxies: ISM --- hydrodynamics --- ISM: clouds --- ISM:
kinematics and dynamics --- stars: formation --- turbulence}


\section{Introduction}
\label{intro}

The disk of the Milky Way contains $\sim 10^9$ $\msun$ of molecular
gas \citep{williams97,bronfman00}, mostly arranged in giant molecular clouds
(GMCs) with typical masses of $\sim 10^6$ $\msun$ and densities
$n_{\rm H} \sim 100$ cm$^{-3}$ \citep{solomon87}. Absent other
support, this gas should collapse on its free-fall time scale, $\tff
\sim 4$ Myr, producing new stars at a rate of roughly $\sim 250$
$\msun$ yr$^{-1}$. However, the observed star formation rate (SFR) in
the Milky Way is only $\sim 3$ $\msun$ yr$^{-1}$ \citep{mckee97}. This
surprisingly low star formation rate, first pointed out by
\citet{zuckerman74}, remains one of the major unsolved riddles for
theories of the interstellar medium (ISM).

In the last 30 years, observations of star formation tracers such as
H$\alpha$ in other galaxies have shown that the problem is not limited
to the Milky Way. \citet{wong02} inferred gas depletion times, defined
as the ratio of the molecular surface density to the star formation
rate per unit area, of a few Gyr in resolved observations of seven
nearby galaxies. This is two orders of magnitude larger than the
typical free-fall times of a few tens of Myr they inferred based on
the cloud densities. \citet{rownd99} and \citet{young96} obtain
similar gas depletion times from unresolved observations in many other 
galaxies. Nor is the problem limited to normal disk galaxies like the Milky
Way. In 87 starbursts \citet{gao04} find CO gas depletion times of
several $0.1 - 1$ Gyr, a factor of ten or less smaller than that in
disk galaxies, and still much longer than typical free-fall
times. \citet{downes98} obtain relatively similar depletion times at
comparable densities for circumnuclear starbursts in 3 nearby
galaxies, and this range of depletion times and characteristic
free-fall times seems typical of starbursts \citep{kennicutt98b}.

An interesting addition to this problem is that the star formation
rate follows clear correlations. Surveys of many galaxies over a range
of star formation rates and surface densities show that the star
formation rate per unit area obeys the Kennicutt-Schmidt Law, which
can be stated in two forms, both equally consistent with observations: 
\begin{equation}
\label{ks1}
\Ssfr \propto \Sg^{1.4}
\end{equation}
or
\begin{equation}
\label{ks2}
\Ssfr \propto \frac{\Sg}{\tdyn},
\end{equation}
where $\Ssfr$ is the star formation rate per unit area, $\Sg$ is
the surface density of gas, and $\tdyn$ is the dynamical (i.e. orbital)
time scale of the galactic disk (Schmidt 1959,1963;
Kennicutt 1998a,1998b; Schmidt's two papers proposed a relationship
between gas density or surface density and star formation rate, while
Kennicutt's determined the exponents and coefficients of the
correlations in equations \ref{ks1} and \ref{ks2} from a large galaxy
sample). Both forms fit the the observed sample of galaxies very well
over a range of nearly eight orders of magnitude in star formation
rate.

Any successful theory of star formation must be able to reproduce both
the lower-than-expected star formation rate and both forms of the
Kennicutt-Schmidt Law, and must do so using physics that is applicable
in a range of environments from Milky Way-like disk galaxies where the
ISM is entirely atomic and the star formation rate is low, to ULIRGs,
where the ISM is fully molecular and the star formation rate is many
orders of magnitude larger. To date, no theory is able to meet these
requirements. Recent numerical work has been able to reproduce some of
the observations, but only with considerable assumptions and
limitations. \citet{kravtsov03} uses the probability distribution of
densities in simulations to suggest that the fraction of high density
gas varies with the overall density to roughly the 1.4 power,
explaining one form of the Kennicutt-Schmidt Law. However, this
observation does not explain the other form of the Law, and it also
says nothing about the absolute rate at which star formation
occurs. It also fails to explain the choice of density cutoff that
constitutes ``high density.'' Similarly, \citet{li05} show that their
simulations reproduce the $\Sg^{1.4}$ form of the Kennicutt-Schmidt
Law. However, their simulations depend on both an arbitrarily density
chosen threshold for star formation and an arbitrary choice of the
star formation rate in gas denser than the threshold.

\citet{tan00} proposes an analytic theory based on star formation
induced by cloud-cloud collisions to explain the Kennicutt-Schmidt
Law. In this model, the star formation rate is proportional to
$\Sg/\tdyn$ because the inter-cloud collision time is proportional to
the dynamical time, and the supply of gas available is proportional to 
the gas surface density. However, this theory also relies on an
unknown efficiency of (collision-induced) star formation that can be
roughly calibrated from observations, but is not independently
predicted. Similarly, \citet{silk97} proposes a theory in which the
star formation rate is set by supernova feedback. However, the theory
depends critically on the porosity $P$ of the interstellar medium to
gas heated by supernovae, and it is unclear how $P$ varies from the
predominantly atomic, diffuse gas disks found in normal galaxies like
the Milky Way to the dense, entirely molecular interstellar media
found in starbursts. In particular, the theory predicts that, if $P$
is roughly constant (as is required to obtain the observed star
formation rate and the Kennicutt-Schmidt Law), then all galaxies
should have the same ISM velocity dispersion. This prediction clearly
fails in starbursts \citep{downes98}.

Another broad class of theories appeals to magnetic fields and ambipolar
diffusion. In these models, star-forming regions are threaded by a
magnetic field strong enough to make them magnetically sub-critical,
so the collapse time is set by the time required for the field to
escape from the gas via ambipolar diffusion (see reviews by Shu et
al. 1987 and Mouschovias 1987; for a more recent discussion, see
Tassis \& Mouschovias 2004). While we discuss this theory in more
detail in \S~\ref{magdiscussion}, we note that observations of magnetic
field strengths in Milky Way GMCs, both directly via Zeeman splitting
\citep{crutcher99, bourke01} and indirectly via statistical indicators
\citep{padoan04b}, suggest that their magnetic fields are not strong
enough by themselves to prevent rapid collapse. Nothing is known of
magnetic field strengths in other galaxies, so it is unknown if this
model can explain the Kennicutt-Schmidt Law.

A final class of theories, on which we shall focus, relies on
turbulence. Observed GMCs in the Milky Way and in nearby galaxies have
significant non-thermal linewidths (e.g. Fukui et al. 2001,
Engargiola et al. 2003, Rosolowsky \& Blitz 2005), and this is
generally interpreted as indicating the presence of supersonic
turbulence. In a cloud supported against collapse by supersonic
turbulence, at any given time most of the mass should be in structures
that are insufficiently dense to collapse (see reviews by Mac Low \&
Klessen 2004 and Elmegreen \& Scalo 2004). This conclusion is
bolstered by simulations (e.g. Klessen, Heitsch, \& Mac Low 2000; Li
et al. 2004) that show that, under at least some circumstances,
supersonic turbulence can inhibit star formation.

\citet{padoan95} provides an analytic theory of the star formation
rate in a turbulent medium that depends on the properties of GMCs, and
on the distribution of masses of clumps that results from
turbulent fragmentation. For Milky Way GMCs it produces a value of the 
star formation rate reasonably in agreement with observations, but
there is no way to extend this result to galaxies where we cannot
directly observe the GMCs. Similarly,
\citet{elmegreen02, elmegreen03} uses the probability
distribution function of densities in a turbulent medium to estimate
the mass fraction of galactic GMCs above a critical density of $\sim
10^5$ cm$^{-3}$, and argue that this can explain the low star
formation rate. However, it is not clear why the critical density is
$10^5$ cm$^{-3}$, or how this value might vary from galaxy to
galaxy. Nor is it clear how this analysis leads to the
Kennicutt-Schmidt Law. Elmegreen argues that the law $\Ssfr \propto
\Sg^{1.4}$ can be explained in this picture if all galaxies have
roughly the same scale height, but does not provide a physical reason
why the scale height should be constant.

Our goal in this paper is to provide a theory of the star formation
rate that can explain both the surprisingly low star formation
rate and two forms of the Kennicutt-Schmidt Law, and that can do so
over a range of conditions from normal disks to ULIRGs. In other words, we
seek to explain both the exponents and the coefficients of the
Kennicutt-Schmidt Laws over their entire observed range. Our theory
does not depend on an unknown efficiency or critical density for star
formation. Instead, we proceed from three premises that are
well-motivated by a combination of observations, simulations, and
theoretical considerations. First, we assume that star formation
occurs primarily in molecular clouds that are virialized and
supersonically turbulent. Second, we assume that the probability
distribution of densities is lognormal, as expected for supersonic
isothermal turbulence. Third, we assume that gas collapses in regions
where the local gravitational potential energy exceeds the local
turbulent energy. In \S~\ref{theory}, we develop these
premises to compute the star formation rate in a cloud in terms of its 
Mach number and virial parameter. We check this theory against
simulations, and show that it is able to reproduce them well.
In \S~\ref{galaxysec}, we apply our estimate to galaxies, and derive
an estimate for the star
formation rate as a function of the observable properties of
galaxies. In \S~\ref{milkywaysec}, we compare our theoretical
predictions to the observed star formation rates in the Milky Way, and 
in \S~\ref{galaxycomp} we compare to a large sample of
galactic-average star formation rates. We show 
that our theory provides an excellent fit to the data. In
\S~\ref{futuretests} we present three future observations that can be
used to check our theory. Finally,
in \S~\ref{discussion} and \S~\ref{conclusion} we discuss and
summarize our conclusions.

\section{The Star Formation Rate Per Free-Fall Time}
\label{theory}

In this section we present our general theory of turbulent regulation
of the star formation rate in dimensionless terms. For convenience we
define the dimensionless star formation rate per free-fall time,
$\sfrff$, which is the fraction of an object's gaseous mass that is
transformed into stars in one free-fall time at the object's mean
density.

\subsection{Derivation}

Both simulations and observations of turbulence in the interstellar
medium show that the turbulent velocity dispersion $\sigma_l$ computed
over a volume of characteristic length $l$ increases with $l$ as
$\sigma_l\propto l^{p}$ with $p\approx 0.5$ \citep{larson81,
solomon87, heyer04}. This self-similar structure appears to be a
universal property of supersonic turbulence, and holds over a very wide
range of length scales in molecular clouds. \citet{ossenkopf02}
summarize observations of
the Polaris Flare molecular cloud that show the linewidth-size
relation over three orders of magnitude in length. Because velocity
dispersions are smaller on smaller scales, even though the velocity
dispersion may be supersonic over length scales comparable to the size
of a simulation box or an entire star-forming cloud, there will be
some smaller scale over which it is not. V\'azquez-Semadeni,
Ballesteros-Paredes, \& Klessen (2003, hereafter VBK03) show that the
scale at which the turbulence transitions from supersonic to subsonic,
the sonic length $\ls$, is a key determinant of whether $\sfrff$ will
be high or low. For the purposes of this paper, we adopt a more
specific definition of the sonic length, consistent with that of
VBK03: let $\sigma_l(\mathbf{x})$ be the one-dimensional velocity
dispersion computed over a sphere of diameter $l$ centered at position
$\mathbf{x}$ within a turbulent medium, and let
\begin{equation}
\sigma_l=\left\langle \sigma_l(\mathbf{x})\right\rangle_{V}
\end{equation}
be the volume average of $\sigma_l(\mathbf{x})$ over the entire
region. We define $\ls$ as the length $l$ such that $\sigma_l=\cs$,
where $\cs$ is the isothermal sound speed in the region. (Note that
our $\ls$ is the same as the turbulent pressure length $l_P$
introduced by Wolfire et al. 2004). The linewidth-size relation then
becomes
\begin{equation}
\label{linewidth1}
\sigma_l=\cs \left(\frac{l}{\ls}\right)^p.
\end{equation}

While VBK03 show that the sonic length correlates well with $\sfrff$,
the star formation rate per free-fall time is a dimensionless number
and the sonic length is a length. On dimensional grounds, there must
therefore be another length scale that is relevant. The natural
candidate is the Jeans length,
\begin{equation}
\lj = \sqrt{\frac{\pi \cs^2}{G\rho}},
\end{equation}
where $\cs$ is the sound speed and $\rho$ is the density at a given
point. Of course in a turbulent medium $\rho$ varies from place to
place, and we account for this effect below. Consider a ``core'', a
sphere of gas embedded in the cloud. The thermal pressure at the
surface of the sphere is roughly $\rho \cs^2$. The largest mass such an
object can have and remain stable against gravitational collapse is
the Bonnor-Ebert mass \citep{ebert55, bonnor56},
\begin{eqnarray}
M_{\rm BE} & = & 1.18 \frac{\cs^3}{\sqrt{G^3 \rho}} \\
& = & \frac{1.18}{\pi^{3/2}} \rho \lj^3.
\end{eqnarray}
The radius of such a sphere is roughly
\begin{equation}
R_{\rm BE} \approx 0.37 \lj.
\end{equation}
The gravitational potential energy of the sphere is
\begin{equation}
\calw = -\frac{3}{5} a \frac{GM_{\rm BE}^2}{R_{\rm BE}}
= -1.06\, \frac{\cs^5}{G^{3/2} \rho^{1/2}}.
\end{equation}
Here $a$ is a geometric factor set by the sphere's mass distribution,
and in the numerical evaluation we have used $a=1.2208$, the value
for a maximum-mass stable Bonnor-Ebert sphere \citep{mckee99b}.
The thermal energy of the gas is
\begin{equation}
\calt_{\rm th} = \frac{3}{2} M_{\rm BE} \cs^2 = 1.14 \,
\left|\calw\right|.
\end{equation}
Using the linewidth-size relation (\ref{linewidth1}), the average
turbulent kinetic energy in the sphere is
\begin{eqnarray}
\calt_{\rm turb} & = & \frac{3}{2} M_{\rm BE} \, \sigma^2\left(2 R_{\rm
BE}\right) \\
& = & 1.14\, \left(0.74\right)^{2p}
\left(\frac{\lj}{\ls}\right)^{2p} \left|\calw\right| \\
& \rightarrow & 0.89 \, \left(\frac{\lj}{\ls}\right)
\left|\calw\right|,
\end{eqnarray}
where for the numerical evaluation in the final step we have used
$p=0.5$.
Thus, a Bonnor-Ebert-mass object has approximately equal kinetic,
thermal, and potential energies if $\lj\sim\ls$. If $\lj\ltsim\ls$,
gravity is approximately balanced by thermal plus turbulent pressure,
and the object is at best marginally stable against collapse. If
$\lj\gg\ls$, kinetic energy greatly exceeds both
potential and thermal energy, and the object is stable against
collapse.

Since $\lj$ is a function of the local density, the condition $\lj
\ltsim \ls$ for collapse translates into a minimum local density
required for collapse. We can use this to compute the star formation
rate, by first estimating what fraction of the mass is at densities
higher than this minimum. Numerous numerical and theoretical studies
find that the probability distribution function (PDF) of the
density in a supersonically turbulent isothermal gas is lognormal,
with a dispersion that increases with Mach number
\citep{vazquezsemadeni94, padoan97, scalo98, passot98, nordlund99,
ostriker99, padoan02}. \citet{padoan02} find that the PDF is
well-fit by the functional form
\begin{equation}
\label{rhopdf}
dp(x) = \frac{1}{\sqrt{2\pi \sr^2}} \exp\left[-\frac{\left(\ln x -
\overline{\ln x}\right)^2}{2\sr^2}\right] \frac{dx}{x}
\end{equation}
where $x=\rho/\rho_0$ is the density normalized to the mean density in 
the region $\rho_0$. The mean of the log of density is
\begin{equation}
\overline{\ln x} = -\frac{\sr^2}{2},
\end{equation}
and the dispersion of the PDF is approximately
\begin{equation}
\label{sreqn}
\sr \approx \left[\ln\left(1+\frac{3 \calm^2}{4}\right)\right]^{1/2},
\end{equation}
where $\calm$ is the one-dimensional Mach number of the turbulent
region measured on its largest scale. Let $\ljzero$ be the Jeans length
at the mean density. The Jeans length at overdensity $x$ is
$\lj(x)=\ljzero/\sqrt{x}$, which we wish to compare to $\ls$. We
therefore define the critical overdensity required for collapse as
\begin{equation}
x \ge \xc \equiv \left(\phix \frac{\ljzero}{\ls}\right)^2,
\end{equation}
where $\phi_x$ is a numerical factor to be determined by fitting in
\S~\ref{simcomparison}. Gas at an overdensity of $\xc$ or higher has a
local Jeans length smaller than the sonic length, and is therefore
unstable to collapse. The fraction of the mass in collapsing
structures is therefore just the fraction of mass at overdensities of
$\xc$ or greater, which is 
\begin{equation}
f = \int_{\xc}^{\infty} x \frac{dp}{dx} \, dx.
\end{equation}

To convert $f$ to a star formation rate, we must account for
two factors. First, approximately $25\%-75\%$ of the mass in
star-forming cores will be ejected by outflows \citep{matzner00}. We
define $\ec$ as the fraction of the mass that reaches the collapsing
core phase that eventually winds up in a star, and adopt a fiducial
value of $\ec=0.5$. Second, we have computed the fraction of mass in
collapsing structures at any given time. To convert this to a rate, we
must divide by the characteristic time scale over which new gas
becomes unstable. When a region collapses, it detaches from the
turbulent flow and thereby removes pressure support from the
remaining, stable gas. The remaining gas will respond to this loss of
pressure support on its gravitational collapse time scale, the
free-fall time. We therefore estimate that new gas becomes
gravitationally unstable over a free-fall timescale
$\tff$. (Alternately, we could have used a crossing time, which is
very similar in a real GMC.)
However, this is just a rough argument, so we let the true time scale
be $\phit \tff$. We will determine $\phit$ for purely hydrodynamic
turbulence in \S~\ref{simcomparison}. Magnetic fields
can delay collapse and make $\phit$ somewhat larger
for a real cloud than our fit will find \citep{vazquezsemadeni05}.

\begin{figure}
\epsfig{file=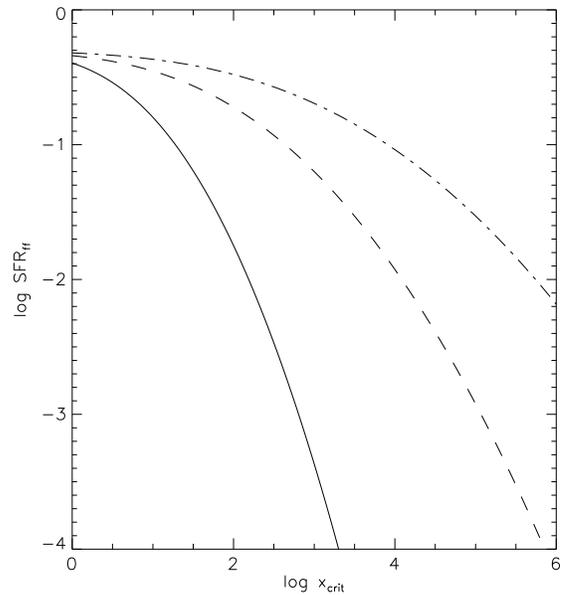}
\caption{\label{SFRplot}
Star formation rate per free-fall time versus critical overdensity,
for Mach numbers of 5 (solid line), 50 (dashed line), and 500
(dot-dashed line).
}
\end{figure}

With these two factors defined, the star
formation rate per free-fall time is
\begin{eqnarray}
\sfrff & = & \frac{\ec}{\phit} \int_{\xc}^{\infty} x p(x) \, dx \\
& = & \frac{\ec}{2 \phit} \left[1+\mbox{erf}\left(\frac{-2 \ln
\xc + \sr^2}{2^{3/2} \sr}\right)\right]
\label{SFRFF}
\end{eqnarray}
The total star formation rate arising from a cloud of mass $M_{\rm
mol}$ is
\begin{equation}
\label{sfrdefn}
\sfr = \sfrff \frac{M_{\rm mol}}{\tff}.
\end{equation}
We plot $\sfrff$ as a function of $\xc$ for $\phit=1$ in Figure
\ref{SFRplot}. We can also define a ``core formation rate'' $\cfrff$,
which reflects the rate at which mass begins to collapse, ignoring
what fraction of it will be ejected by feedback. This is simply $\sfrff$
with $\ec=1$.

\citet{padoan95} and \citet{padoan02,padoan04} have
previously approached the problem of estimating the star formation
rate by considering the combination of the PDF of densities and the
mass distribution of clumps created by fragmentation in a turbulent
medium. Since we are interested only in the rate at which stars form,
and not their mass distribution, we may neglect the clump mass
distribution and consider only the distribution of densities. In so
doing, we implicitly assume that all or most of the mass that is at
densities rendering it capable of collapse will in be in the presence
of enough other high-density gas so that it does collapse. This
assumption is bolstered by the observation that turbulence tends to
organize mass into filaments and voids on large scales, so that high
density gas is likely to be in the presence of other high density
gas. Moreover, as we show in \S~\ref{simcomparison}, this assumption
produces a theory that shows good agreement with simulations.

\subsection{Comparison to Simulations}
\label{simcomparison}

To test our theory, we compare to the work of VBK03, who simulated a
turbulent periodic box of gas and computed the fraction of the mass
that collapsed into stars. The simulation setup is described in detail in
\citet{klessen00}, but we summarize it here. In simulation units, the
box length is $L=2$, the sound speed $\cs=0.1$, the mean density is
$\rho_0=1/8$, the Jeans length at that density is $\ljzero=1/2$, and
the free-fall time is $\tff=1.5$. Turbulence is driven at a
one-dimensional Mach number $\calm=2$, $3.2$, $6$, or $10$ using a
driving field that contains power only at wavenumbers around $k=2$,
$4$, or $8$, where $k \equiv L/\lambda_d$ and $\lambda_d$ is the
driving wavelength.

We read off the sonic length from Figure 3c of VBK03, noting that
VBK03 define the sonic length using the three-dimensional velocity
dispersion, while we use the one-dimensional velocity
dispersion (J. Ballesteros-Paredes and E. V\'azquez-Semadeni, 
private communication). Given the scaling $\sigma\propto l^p$, and
assuming the turbulent velocity field is roughly isotropic, the two
are related by $\ls \approx 3^{1/(2p)} \lsthree$. We adopt $p=0.5$
through the rest of this section. To determine the $\sfrff$, we read
off data from Figure 2 of VBK03. We measure the time $t$ at which a
fraction $f=0.1$ of the mass in the run has collapsed into stars. For
runs where less than $10\%$ of the mass has collapsed by the end, we
measure $f$ and $t$ at the point where the run ends. We then compute
$\sfrff=1.5 \,f/t$. (Using $20\%$ instead of $10\%$ did not
substantially change the result.) We summarize all of this in Table
\ref{VBKtab}.

\begin{table}
\centerline{
\begin{tabular}{ccccccc}
\hline\hline
Run name &
$\lsthree$ &
$t$ &
$\mbox{SFR}_{\rm ff-sim}$ &
$\mbox{SFR}_{\rm ff-th}$ \\ \hline
M2K2	& 0.16	& 0.62	& 0.24	& 0.33\\
M2K4	& 0.10	& 0.62	& 0.24	& 0.18\\
M2K8	& 0.20	& 0.62	& 0.24	& 0.39\\
M3.2K2	& 0.080	& 0.44	& 0.34	& 0.18\\
M3.2K4	& 0.046	& 1.58	& 0.095	& 0.0641\\
M3.2K8	& 0.031	& 2.48	& 0.060	& 0.023\\
M6K2	& 0.039	& 0.30	& 0.50	& 0.11\\
M6K4	& 0.023	& 2.27	& 0.066	& 0.045\\
M6K8	& 0.016	& 6.89	& 0.022	& 0.019\\
M10K2	& 0.018	& 0.87	& 0.17	& 0.060\\
M10K4
\footnote{Run ended with $f=0.058$ of mass collapsed}
	& 0.013	& 6.03	& 0.014	& 0.035\\
M10K8
\footnote{Run ended with $f=0.084$ of mass collapsed}
	& 0.0094	& 4.69	& 0.026	& 0.018
\\
\hline\hline
\end{tabular}
}
\caption{\label{VBKtab}
Col. (1): Run name in VBK03. M$m$K$k$ indicates that
the 1-D Mach number is $m$ and the run is driven at wavenumber
$k$. Col. (2): Measured value of $\lsthree$. Col. (3): Time at which
$10\%$ of the mass had collapsed, or when the run ended, in code
units. Col. (4): Star formation rate in the simulation, defined
as $\sfrff=f/(t/\tff)$. Col. (5): Theoretically estimated $\sfrff$.
}
\end{table}

\begin{figure}
\epsfig{file=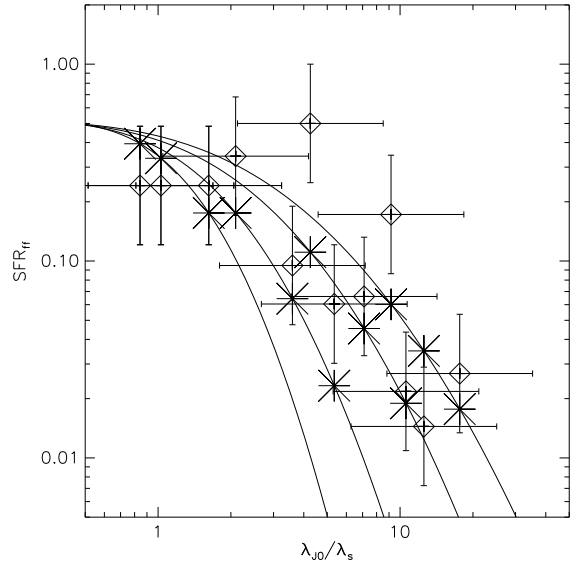}
\caption{\label{SFRsim}
Star formation rate per free-fall time versus $\ljzero/\ls$, as
measured from the VBK03 runs (error bars with diamonds) and as
predicted by our theoretical model (asterisks). The lines show
our theoretical predictions of $\sfrff$ versus $\ljzero/\ls$ for Mach
numbers of 2 (lowest line), 3.2 (second line), 6 (third line), and 10
(highest line).
}
\end{figure}

We fit the VBK03 data to our theoretical estimate of $\cfrff$
rather than $\sfrff$ because the VBK03 simulations do not include any
feedback. The cases with large $\xc$ are closest to the environment in
real star-forming clouds, so we weight by $\xc^2$. A
Levenberg-Marquardt fit with this weighting gives $\phix = 1.12$ and
$\phit = 1.91$. We compare the simulation to $\cfrff$ evaluated with
the best-fit values in Table \ref{VBKtab} and in Figure
\ref{SFRsim}. In the Figure, the simulation points have error bars
corresponding to a factor of 2 uncertainty, as recommended by
VBK03. As the plot shows, there is a large scatter, but we are able to
reproduce the overall behavior of the VBK03 simulation data quite
well. Note that $\phit = 1.91$ implies that, for virialized objects,
the characteristic time scale is roughly a single crossing time -- see 
\S~\ref{turbdecay}.

To understand how magnetic fields might change our results, we examine
the work of \citet{li04}, who measure the amount of mass collapsed
into cores as a function of time a magnetohydrodynamic periodic box
simulation similar to those of VBK03 (identical box length, Jeans
length, sound speed, and free-fall time). The initial box is
magnetically supercritical, with $M/\mphi=8.3$. The simulation
is driven with a three-dimensional Mach number of 10 (one-dimensional
Mach number $\calm=5.8$) at a driving wavenumber of $k=2$, and is
therefore very similar to run M6K2 in VBK03. \citet{li04} do not
measure a sonic length, so we use the measured sonic length of
$\lsthree=0.039$ from the corresponding VBK03 run. With these
parameters and our best-fit values of $\phix$ and $\phit$, we find
$\cfrff=0.11$. Reading off the time at which $10\%$ of the mass had
collapsed in the highest resolution run from Figure 6 of \citet{li04}
gives a measured $\cfrff=0.072$. The simulation result is slightly
lower, but well within the factor of two error recommended by
VBK03. While this is only one simulation, it provides some confidence
that the inclusion of magnetic fields in the supercritical regime will
not change the star formation rate substantially.

\subsection{$\sfrff$ in Virialized Objects}
\label{virialobj}

Using our theory, we can compute $\sfrff$ in virialized molecular
clouds and clumps. \citet{bertoldi92} define the virial parameter for
a spherical cloud as
\begin{equation}
\label{alphadef}
\avir = \frac{5\sigma_{\rm tot}^2 R}{GM},
\end{equation}
where $\sigma_{\rm tot}$ is the one-dimensional thermal plus turbulent
velocity dispersion over the entire cloud, $R$ is the radius of the
cloud, and $M$ is the mass. Since we are concerned with large
star-forming clouds that have $\sigma_{\rm tot}\gg \cs$,
$\sigma_{\rm tot}$ is approximately equal to the turbulent
velocity on the largest scale, which we denote $\sigma_{2R}$. Clouds with
$\avir\approx 1$ are in self-gravitating virial equilibrium, meaning
that internal pressure (turbulent plus thermal) approximately
balances gravity. Clouds with $\avir \gg 1$ are non-self-gravitating
and are confined by external pressure, while $\avir \ll 1$ indicates
either that the cloud is supported against gravity by a magnetic
pressure larger than either the turbulent or thermal pressure, or that
the cloud is undergoing free-fall collapse. We refer to objects with
$\avir\approx 1$ as ``virialized.''

Consider now a star-forming region that follows the linewidth-size
relation
\begin{equation}
\label{linewidth2}
\sigma_l = \sigma_{2R} \left(\frac{l}{2R}\right)^{p}.
\end{equation}
The sonic scale is therefore
\begin{equation}
\ls = 2R \left(\frac{\cs}{\sigma_{2R}}\right)^{1/p},
\end{equation}
and the Jeans length at the mean density $\rho_0$ is
\begin{equation}
\ljzero = \sqrt{\frac{\pi \cs^2}{G \rho_0}}
= 2\pi \cs \sqrt{\frac{R^3}{3 G M}}.
\end{equation}
Thus, the critical overdensity required for collapse is
\begin{eqnarray}
\xc & = & \left(\phix \frac{\ljzero}{\ls}\right)^2 \\
& = & \frac{\pi^2 \phix^2}{15} \avir
\calm^{\frac{2}{p}-2}
\label{xcrit}
\\
& \rightarrow & 1.07 \calm^2,
\end{eqnarray}
where $\calm=\sigma_{2R}/\cs$ is the Mach number of the region. We
have used the definition of the virial parameter
(\ref{alphadef}) in the second step, and for the numerical evaluation
we have used our best-fit value of $\phix$ and taken $\avir=1.3$. This 
choice is based on the evaluation of Milky Way GMCs performed by
\citet{mckee03}. We discuss it in more detail in
\S~\ref{bounddiscussion}. From this formulation, it is
straightforward using (\ref{SFRFF}) to compute $\sfrff$ in a cloud in
terms of $\avir$ and $\calm$ for the cloud. We have therefore
succeeded in computing the dimensionless star formation rate $\sfrff$
in terms of the two basic dimensionless numbers that describe a
turbulent cloud: the ratio of kinetic to potential energy (roughly
$\avir$) and the ratio of kinetic to thermal energy (roughly
$\calm^2$). This relation has an intuitive physical interpretation.
At an overdensity of $\xc$, the thermal pressure is
\begin{equation}
P_{\rm th} = \rho \cs^2 \approx \rho_0 \sigma_{2R}^2 = P_{\rm turb}.
\end{equation}
Thus, the gas capable of collapse is simply the gas that is dense
enough so that its thermal pressure is comparable to or greater than
the mean turbulent pressure in the cloud, $P_{\rm turb}$

\begin{figure}
\epsfig{file=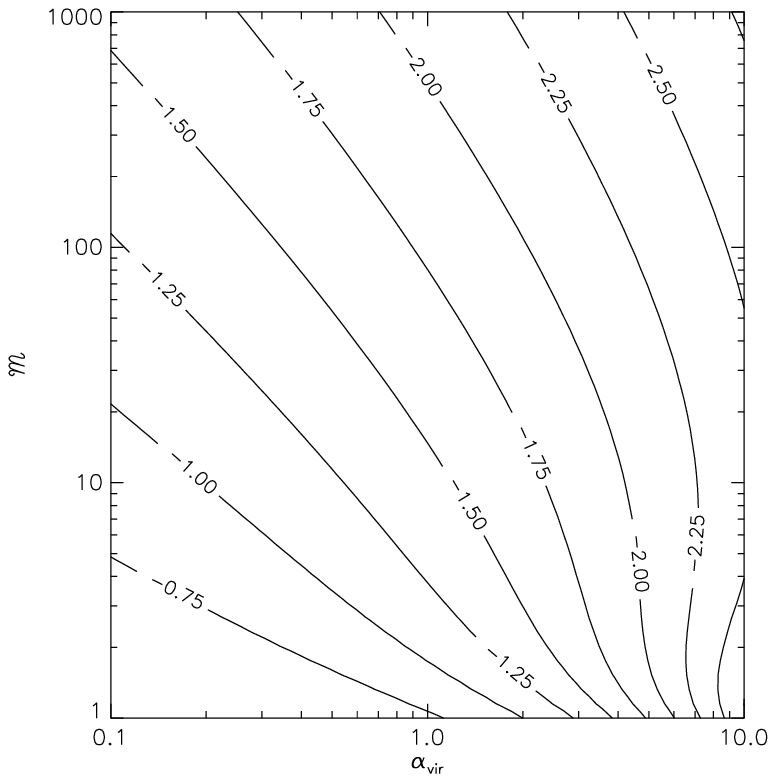}
\caption{\label{sfrvir}
Contours of star formation rate per free-fall time $\sfrff$ versus
$\avir$ and $\calm$. The contours are labelled by value of
$\log\sfrff$.
}
\end{figure}

\begin{figure}
\epsfig{file=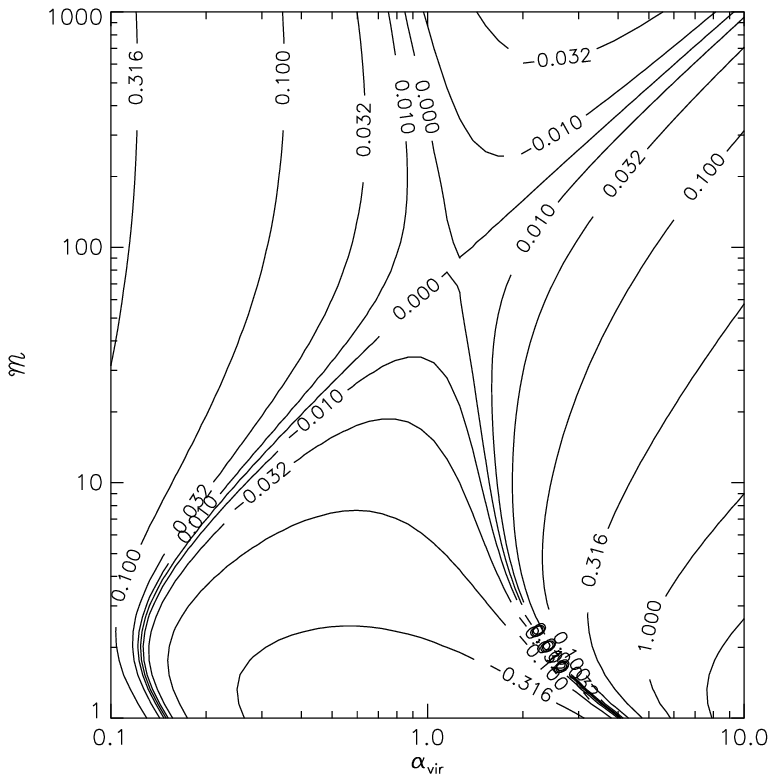}
\caption{\label{sfrvirfit}
Contours of the error in our power-law fit for $\sfrff$, defined as
$\mbox{error} = (\mbox{fit} - \sfrff)/\sfrff$. The labels on the
contours show the value of the error.
}
\end{figure}

In Figure \ref{sfrvir} we plot the star formation rate per free-fall
time as a function of $\avir$ and $\calm$ for $p=0.5$. For
convenience, we also fit $\sfrff$ by a power law,
\begin{equation}
\label{sfrfffit}
\sfrff\approx 0.014
\left(\frac{\avir}{1.3}\right)^{-0.68}
\left(\frac{\calm}{100}\right)^{-0.32}.
\end{equation}
Figure \ref{sfrvirfit} shows the error in our power-law fit as a
function of $\calm$ and $\avir$. The error is less than $5\%$ for
values of $\avir$ from $\sim 0.5-3$ and $\calm$ from $\sim
10-1000$. Since real star-forming clouds generally
fall within this range (see \S~\ref{galaxysec}), this power law is a
reasonably good approximation. One important thing to note about
$\sfrff$ is how weakly $\sfrff$ varies with $\calm$. Thus, \textit{the
star formation rate per free-fall time in a virialized cloud depends
very weakly on the Mach number of the cloud}. This is easy to
understand intuitively. At fixed
$\avir$, increasing $\calm$ increases $\xc$, raising the overdensity
that the gas must reach to collapse. At the same time, however,
increasing $\calm$ increases the width of the probability distribution 
function, putting a larger fraction of the gas at high
overdensities. These two effects nearly cancel out, which is why
changing $\calm$ at fixed $\avir$ has little effect on $\sfrff$.

Before proceeding, we must point out one limit of our analysis. We
have assumed that the internal structure of GMCs follows the
linewidth-size relation. However, the OB star-forming clumps
observed in CS by \citet{plume97} do not. They have substantially
higher velocity dispersions than is typical for an object of their
size in their parent GMCs, and their sizes and velocity dispersions do 
not appear to be correlated. We interpret these clumps as regions of a
GMC larger than a single core that have become gravitationally
unstable and collapsed to higher surface densities and pressures than
the rest of the GMC \citep{mckee03}, increasing their velocity
dispersions. The VBK03 simulations that we have used to calibrate our
model do not have enough dynamic range to include the presence of such
regions, so our estimate of $\sfrff$ ignores their presence.
Fortunately, clumps of this sort constitute only a tiny
fraction of the total molecular mass of the galaxy, and are even a
small fraction of the mass of their parent GMCs. Thus, the error we
have made by ignoring them is negligible on the large scales with
which we are concerned.

\section{Star Formation in Galaxies}
\label{galaxysec}

In this section we
will usually give surface densities in units of $\msun$ pc$^{-2}$. For
convenience, we note that $1\;\msun\;\mbox{pc}^{-2} = 2.1\times
10^{-4}\mbox{ g cm}^{-2}=8.9\times 10^{19}\mbox{ Hydrogen nuclei
cm}^{-2}$, and $1\;\msun\;\mbox{pc}^{-2}$ corresponds to $A_V=0.045$
for the local dust to gas ratio.

\subsection{The Star Formation Law for Galactic Disks}

Our formulation applies equally well to galactic disks. The star
formation rate per unit area of a galactic disk is simply
\begin{equation}
\Ssfr = \frac{\sfrff \fGMC \Sg}{\tff} \approx
\frac{0.061}{\avir^{0.68}}
\left(\frac{\fGMC \Sg}{\calm^{0.32} \tff}\right)
\end{equation}
where $\Sg$ is the gas surface density of the disk, $\fGMC$ is the
fraction of it that is in molecular clouds,
and $\tff$ and $\calm$ are the characteristic free-fall times and Mach
numbers in the star-forming regions of the disk. To estimate these
quantities, we begin by considering the mean properties of galactic
disks. Note that for galaxies like the Milky Way, essentially all the
molecular gas is in GMCs, so $\fGMC$ is just the molecular
fraction. For starbursts, we also assume that all the molecular gas is
collected into bound clouds, although this is approximate, as we
discuss further in \S~\ref{uncertainties}.

Consider star formation in a galactic disk with a
total surface density of $\Stot$. The pressure at the disk
midplane is then given by (cf. Elmegreen 1989 and Blitz \& Rosolowsky
2004)
\begin{equation}
\label{Pmp}
P_{\rm mp} = \phimp \frac{\pi}{2} G \Sg \Stot = \phimp \fg^{-1}
\frac{\pi}{2} G \Sg^2 \equiv \phip \frac{\pi}{2} G \Sg^2
\end{equation}
where $\phimp$ and $\phip$ are constants of order unity and
$\fg=\Sg/\Stot$ is the gas fraction in the galaxy. For an isothermal
disk consisting entirely of gas, $\fg=\phimp=\phip=1$ exactly. For a
real galactic disk containing stars, $\phip>1$, because the gravity of 
the stars compresses the gas. We show in Appendix \ref{phipapp} that
$\phip\approx 3$. The scale height $\hg$ of the gas in the disk is
related to its midplane density by
\begin{equation}
\label{hgeqn1}
\hg = \frac{\Sg}{2\rhog} = \frac{\sg}{\sqrt{2\pi G \phip \rhog}},
\end{equation}
where $\sg$ is the gas velocity dispersion. Using these two
expressions to solve for the midplane density gives
\begin{equation}
\label{rhogeqn1}
\rhog = \frac{\pi G \phip \Sg^2}{2\sg^2}.
\end{equation}

To use this result, we must know $\sg$, which varies from $\sim 6$ km
s$^{-1}$ in normal disks \citep{blitz04} to
$\sim 100$ km s$^{-1}$ in starbursts (e.g. Downes \&
Solomon 1998). To estimate the velocity dispersion, we assume that the
star-forming part of a galaxy has a flat rotation curve with
velocity $\vrot$ and is marginally Toomre stable, so that
$Q\approx 1$. Both assumptions are well-satisfied in observed
galaxies ranging from normal disks to starbursts, and are expected on
theoretical grounds \citep{quirk72, kennicutt89, navarro97, downes98,
martin01, seljak02, navarro03}. The Toomre parameter $Q$
is defined as \citep{toomre64}
\begin{equation}
\label{Qdef}
Q \equiv \frac{\kappa\sg}{\pi G \Sg} = \frac{\sqrt{2} \Omega \sg}{\pi G
\Sg},
\end{equation}
where $\kappa\approx \sqrt{2} \Omega$ is the epicyclic frequency,
$\Omega = \vrot/r$ is the angular velocity, and $r$ is the
galactocentric radius.We adopt $Q=1.5$ as a typical 
value based on the surveys of \citet{martin01} and
\citet{wong02}. However, observed $Q$ values range from $\sim 0.5$ up
to $\sim 6$, and spiral arms generally decrease $Q$. We discuss the
resulting uncertainty in the star formation rate in
\S~\ref{uncertainties}.

Using (\ref{Qdef}) to eliminate $\sg$ in (\ref{rhogeqn1}), we obtain
the mean density in a galactic disk midplane \citep{thompson05},
\begin{eqnarray}
\label{rhogeqn2}
\rhog & = & \frac{\phip \Omega^2}{\pi Q^2 G} \\
& \rightarrow &
6.4\times 10^{-21} \;
\Qonefive^{-2} \Omzero^2
\mbox{ g cm}^{-3},
\end{eqnarray}
where $\Omzero$ is $\Omega$ in units of $\mbox{Myr}^{-1}$, and
$\Qonefive=Q/1.5$ The corresponding free-fall time in the midplane gas 
is
\begin{eqnarray}
\tffg & = & \left(\frac{3\pi^2}{32}\right)^{1/2} \phip^{-1/2}
\frac{Q}{\Omega} \\
& \rightarrow &
0.83 \; \Qonefive \Omzero^{-1}\mbox{ Myr}.
\end{eqnarray}
Since the filling factor of molecular clouds is less than unity even
in galaxies where the ISM is wholly molecular \citep{rosolowsky05},
the mean gas density in the star-forming clouds will be higher than
this and the free-fall time lower. Let $\phirho$ be the ratio of the
mean molecular cloud density to the mean midplane
density,
\begin{equation}
\phirho \equiv \frac{\rhocl}{\rhog}.
\end{equation}
With this definition, we can write the total star formation
rate as
\begin{eqnarray}
\label{sfrapprox1}
\Ssfr & = & \left(\frac{32}{3\pi^2}\right)^{1/2}
\phip^{1/2} \phirho^{1/2} \sfrff Q^{-1} \fGMC \Sg \Omega \\
& \approx & 
0.073 \; \calm^{-0.32} \phirho^{1/2}
\Qonefive^{-1} \fGMC \Sg \Omega,
\end{eqnarray}
where the numerical evaluation uses our fiducial values of $\avir$ and
$\phip$. Noting that $\Omega\propto \tdyn^{-1}$, we see that our
formulation already gives us something 
like the Kennicutt-Schmidt Law, equation (\ref{ks2}). The Milky Way
values for the remaining parameters are $\calm\approx 25$
\citep{solomon87}, $\phirho\approx 20$ \citep{mckee99},
$\Qonefive\approx 1$, and $\fGMC\approx 0.25$ \citep{dame87}, which
gives a numerical coefficient of $0.03$ in equation
(\ref{sfrapprox1}), within a factor of 2 of the coefficient of $0.017$
determined by \citet{kennicutt98a} based on a large sample of
galaxies. (Note that for the observational value of $\phirho$ we are
comparing the density in GMCs to the density in the spiral arms, which 
is a factor of $\sim 4$ higher than the mean ISM density -- see
Nakanishi \& Sofue 2003.) Thus, our 
theory seems consistent with the observed Kennicutt-Schmidt
Law. However, our results depend on two quantities, $\phirho$ and
$\calm$, that have been directly observed only in the Milky Way and a
few nearby galaxies. To completely derive a star formation law
in terms of observables, we must compute $\phirho$ and $\calm$ in
terms of other quantities. Fortunately, $\phirho$ and $\calm$ enter
our prediction to the $0.5$ and $0.32$ powers, so we are relatively
insensitive to errors in them.

\subsection{The Properties of Molecular Clouds}

Our goal in this section is to estimate $\phirho$ and $\calm$ in terms
of observables. Our strategy is to treat molecular clouds as
gravitationally bound fragments of the interstellar medium in
approximate virial balance. The assumption of gravitational
boundedness allows us the estimate the typical mass of GMCs, and this
mass plus the assumption of virial balance allows us to compute the
overdensity and velocity dispersion in GMCs.

In the Milky Way, most molecular gas is in clouds with
masses of a few $\times$ $10^6$ $\msun$ \citep{solomon87, heyer01},
and the LMC
shows a similar characteristic mass \citep{fukui01}. The typical mass
is somewhat lower in M33 \citep{engargiola03} and higher in M64
\citep{rosolowsky05}, indicating a very rough trend of increasing GMC
mass with increasing galaxy surface density. However, all of these are
ordinary disk galaxies, with surface densities $\ltsim 100$ $\msun$
pc$^{-2}$. There are no observations that resolve individual molecular 
clouds in starbursts or ULIRGs, so we must estimate. Since GMCs appear
to be gravitationally bound, they must have formed via a gravitational
collapse. For this reason, their typical mass should be roughly the
Jeans mass in a galactic disk \citep{kim01, kim02, kim03}, giving
\begin{eqnarray}
\label{jeansmass}
\mcl & \approx & \frac{\sg^4}{G^2 \Sg} \\
& = &
\frac{\pi^4 G^2 \Sg^3 Q^4}{4 \Omega^4} \\
& \rightarrow & 2.5 \times 10^{3} \;
\Qonefive^4 \,
\Sgtwo^3 \,
\Omzero^{-4}
\quad \msun,
\end{eqnarray}
where in the second step we have used the definition of $Q$ (equation
\ref{Qdef}) to eliminate $\sg$, and $\Sgtwo$ is $\Sg$ in units of
$10^2$ $\msun$ pc$^{-2}$.
In the Milky Way near the solar circle, where $\sg\approx 6$
km s$^{-1}$ (consistent with the sound speed in the warm ISM -- see
Heiles \& Troland 2003) and $\Sg\approx 12$ $\msun$ pc$^{-2}$
\citep{boulares90}, equation (\ref{jeansmass}) gives $M_J \approx
6\times 10^6$ $\msun$. This agrees well with observed masses of giant
atomic-molecular complexes, of which GMCs are the inner parts
\citep{elmegreen89, elmegreen94}. Note that the Toomre mass and the
Jeans mass are roughly equal for a disk with $Q\approx 1$. The Toomre
mass is $M_T \sim \lambda_T^2 \Sg$, where $\lambda_T \approx Q\hg$ is
the most unstable wavelength and $\hg$ is the gas scale height. The
Jeans mass is $M_J\sim \hg^2 \Sg$, so $M_T \sim Q^2 M_J$.

Now that we have estimated the typical masses of star-forming clouds,
we can compute their typical densities from knowledge of the pressures
that confine them. The pressure at the surface of a GMC is roughly the
ambient pressure in the midplane of a galaxy, $P_{\rm mp}$.
We define $\phipbar$ as the ratio of the mean pressure in a cloud
$\pcl$ to the surface pressure, so
\begin{equation}
\pcl \equiv \phipbar P_{\rm mp}.
\end{equation}
In an environment with a purely molecular ISM, this is just the ratio
of the mean pressure in a gravitationally bound object to its edge
pressure, and is $\sim 2$. In a predominantly atomic ISM,
$\phipbar$ is larger because molecular gas exists only when it is
shielded by an atomic layer, and the weight of the bound atomic gas
increases its pressure. We estimate $\phipbar\approx 10-8\fGMC$, where
$\fGMC$ is the molecular gas fraction of the galaxy, in Appendix
\ref{phipbarapp}.

We now write down the virial theorem for a GMC, using a form of the
theorem obtained by combining equation (24) of \citet{mckee99} with
equation (A7) of \citet{mckee03}:
\begin{equation}
\label{virialthm}
\pcl = \frac{3\pi}{20} \avir G \Scl^2,
\end{equation}
where $\Scl$ is the surface density of the GMC, and $\avir$ is the
standard virial parameter,
\begin{equation}
\label{alphadef2}
\avir = \frac{5\scl^2 \rcl}{G \mcl} = \frac{5\scl^2}{G \sqrt{\pi \mcl
\Scl}}.
\end{equation}
Equation (\ref{virialthm}) is quite intuitive, as it simply equates
the GMC's internal pressure with its weight, scaled by the virial
parameter as an indicator of how
self-gravitating the cloud is. Together with the definition of the
turbulent pressure $\pcl=\rhocl \scl^2$, (\ref{virialthm}) and
(\ref{alphadef2}) constitute three equations in the unknowns $\rhocl$,
$\scl$, and $\Scl$. Solving for the molecular cloud density gives
\begin{equation}
\rhocl = \left(\frac{375}{4\pi}\right)^{1/4}
\left(\frac{\pcl^3}{\avir^3 G^3 \mcl^2}\right)^{1/4},
\end{equation}
and plugging in for $\pcl$ and $\mcl$ gives
\begin{eqnarray}
\phirho & = & \frac{\rhocl}{\rhog} =
\left(\frac{375}{2\pi^2}\right)^{1/4} 
\left(\frac{\phipbar^3}{\phip \avir^3}\right)^{1/4} \\
& \rightarrow &
5.0 \, \phipbarsix^{3/4},
\end{eqnarray}
where $\phipbarsix\equiv \phipbar/6$. The GMC velocity dispersion is
\begin{eqnarray}
\label{cloudsigma}
\scl & = &
\frac{\pi}{\sqrt{2}}
\sqrt{\frac{\phipbar Q^2}{\phirho}}
\frac{G \Sg}{\Omega}
\\
& \rightarrow &
1.6 \,
\phipbarsix^{1/8} \,
\Qonefive \,
\Omzero^{-1} \,
\Sgtwo
\mbox{ km s}^{-1}.
\end{eqnarray}
The numerical evaluations are for $\phip=3$ and $\avir=1.3$. The range
of variation of $\phirho$ with $\fGMC$ is from $\phirho=7.3$ for
$\fGMC=0$ to $\phirho=2.2$ for $\fGMC=1$. Thus, $\phirho$ is $3-4$
times larger in normal galaxies than in starbursts.

To convert the velocity dispersion (\ref{cloudsigma}) to a Mach
number, we must know the sound speed in the star-forming
clouds. Observations of a galaxy can generally determine
the temperature $T$ in the star-forming gas, from which one can easily
compute the sound speed $\cs=\sqrt{k_B T/m}$, where $m=3.9\times
10^{-24}$ g is the mean particle mass, corresponding to a fully
molecular gas with a ratio of 10 H nuclei per He nucleus. However, for
the purposes of numerical evaluation we can use an average sound speed.
In the Milky Way, the typical temperature in star-forming clouds is
$\sim 10$ K \citep{solomon87}, giving a sound speed of 0.19 km
s$^{-1}$. Observed starbursts have temperatures in the range $29-46$ K
\citep{gao04}, giving sound speeds up to 0.4 km s$^{-1}$. For the
numerical evaluations in this paper we adopt an intermediate value of 
0.3 km s$^{-1}$, although $\cs$ is generally directly observable.
Since the Mach number affects the star formation
rate only through $\sfrff$, and $\sfrff$ is very insensitive to Mach
number, this produces relatively little error. We therefore estimate
the typical Mach numbers in star-forming regions to be
\begin{eqnarray}
\label{macheqn}
\calm & = &
\frac{\pi}{\sqrt{2}}
\sqrt{\frac{\phipbar Q^2}{\phirho}}
\frac{G \Sg}{\cs \Omega}
\\
& \rightarrow & 
5.3 \,
\phipbarsix^{1/8} \,
\Qonefive \,
\Omzero^{-1} \,
\Sgtwo.
\end{eqnarray}
Note that while $\scl$ is actually the total thermal plus non-thermal
velocity dispersion, star-forming regions are highly supersonic, so
$\scl \approx \sigma_{\rm non-thermal}$.

\subsection{The Full Star Formation Rate}

Using our calculated values for $\phirho$ and $\calm$, the star
formation rate per unit area of a galactic disk is 
\begin{eqnarray}
\label{sfr1}
\Ssfr 
& = &
\frac{2^{19/8} 5^{3/8}}{3^{3/8} \pi^{5/4}}
\left(\frac{\phip \phipbar}{\avir}\right)^{3/8} Q^{-1}
\sfrff \fGMC \Omega \Sg \\
& \rightarrow &
\label{sfrfit}
9.5 \;
\fGMC \,
\phipbarsix^{0.34}
\Qonefive^{-1.32} \,
\Omzero^{1.32} \,
\Sgtwo^{0.68}
\nonumber \\
& & \qquad
\msun\mbox{ yr}^{-1}\mbox{ kpc}^{-2},
\end{eqnarray}
where the numerical evaluation uses our power law fit for $\sfrff$
(equation \ref{sfrfffit}) and the fiducial values of
all our other parameters, as summarized in Table 
\ref{paramlist}. If one uses our approximation for $\phipbar$ in terms 
of $\fGMC$, this formulation of the star formation rate
now depends solely on observables. Note that our result is
different than the standard scalings with $\Sg$ and $\Omega$ found by
\citet{kennicutt98a}, and it is therefore a new prediction that can be
tested against future observations. Also note that this relation
should apply not just on a galaxy-by-galaxy basis, but within an
individual galaxy as well. This t0o is a new observational
prediction. We discuss ways of testing these predictions in
\S~\ref{futuretests}.

\begin{table}
\centerline{
\begin{tabular}{lr}
\hline\hline
Parameter &
Value \\ \hline
$\avir$	& 1.3 \\
$\cs$	& $0.3$ km s$^{-1}$ \\
$\ec$	& 0.5\\
$p$	& 0.5 \\
$\phip$	& 3.0 \\
$\phipbar$ & $10-8\fGMC$ \\
$\phit$	& 1.91 \\
$\phix$	& 1.12 \\
$Q$	& 1.5
\\ \hline\hline
\end{tabular}
}
\caption{\label{paramlist}
Col. (1): Parameter. Col. (2): Adopted value.
}
\end{table}

\section{Comparison to the Milky Way}
\label{milkywaysec}

We first test our theoretical prediction against the Milky Way. We do
so in two ways to show that the our results are consistent. First we use
the observed properties of the molecular gas in the Milky Way plus our
estimate of $\sfrff$, and second we use the estimated surface
densities of various MW components.

\subsection{Estimate From Observed GMC Properties}
\label{gmcpropsest}

\citet{bronfman00} estimate that the total mass of GMCs inside the
solar circle is $M_{\rm mol} \approx 10^9$ $\msun$. The mass
distribution of the clouds is \citep{williams97}
\begin{equation}
\frac{d\mathcal{N}}{d\,\ln M_{\rm cl}} \approx
\left\{
\begin{array}{lr}
0, & \qquad M_{\rm cl,6} > 6 \\
10 \,M_{\rm cl,6}^{-0.6},
&
\qquad M_{\rm cl,6}< 6
\end{array}
\right.
,
\end{equation}
where $M_{\rm cl}$ is the cloud mass and $M_{\rm cl,6}=M_{\rm
cl}/(10^6\;\msun)$.
\citet{solomon87} catalog 273 galactic GMCs observed in CO
They find that the average column density of GMCs
is $N_{\rm H} \approx  1.5 \times 10^{22}$ cm$^{-2}$ independent of
mass, where the subscript H indicates that we  are referring to the
number of hydrogen nuclei. \citet{mckee99} uses this result to
estimate that the free-fall time in a GMC is
\begin{equation}
\tff = 4.7 \left(\frac{M_{\rm cl}}{10^6\;\msun}\right)^{1/4}
\mbox{ Myr}.
\end{equation}
Combining the linewidth-size and mass-radius relations inferred by
\citet{solomon87} and \citet{mckee99} gives a Mach number-radius
relation
\begin{equation}
\calm_{\rm cl} = 25 \left(\frac{M_{\rm
cl}}{10^6\;\msun}\right)^{0.25}
\end{equation}
for a GMC temperature of 10 K.
From these relations, it is straightforward to estimate the total star 
formation rate by integrating the star formation rate over the GMC
mass distribution,
\begin{eqnarray}
\label{sfrmw}
\dot{M}_{\rm *-pred}
& = &
\int_{10^4\;\mbox{\scriptsize M$_{\odot}$}}^{6\times
10^6\;\mbox{\scriptsize M$_{\odot}$}}
\frac{
\sfrff(M_{\rm cl})
}{\tff(M_{\rm cl})} \,
\frac{d\mathcal{N}}{d\, \ln M_{\rm cl}} \, dM_{\rm cl} \\
& \approx & 5.3 \;\msun\mbox{ yr}^{-1}.
\end{eqnarray}
We have imposed a lower cutoff of $10^4$ $\msun$ because $\avir\gg 1$
for GMCs with smaller masses \citep{heyer01}, which
greatly reduces their star formation rate. The observed star formation
rate in the Milky Way
is $\dot{M}_{\rm *} \approx 3$ $\msun$ yr$^{-1}$ \citep{mckee97}, so
our estimate agrees with observations to a factor of $1.8$, a
reasonable fit.

An important subtlety of this analysis is that we must impose a lower
cutoff when integrating the star formation rate over the GMC mass
distribution because small clouds, if they are virialized, contribute
significantly to the star formation rate. The integrand in
(\ref{sfrmw}) scales as roughly $M_{\rm cl}^{-0.93}$: one gets an exponent of
$-0.6$ from the logarithmic mass spectrum $d\mathcal{N}/d\ln M_{\rm cl}$,
$-0.25$ from the free-fall time, and $\sim -0.08$ from
the dependence of $\sfrff$ on the Mach number, and hence on $M_{\rm
cl}$. Thus, each decade range in the mass of virialized clouds
contributes almost equally to the star formation rate. The
contribution to the total star formation rate from small clouds is
small not because the clouds contain a small amount of mass, but
because small clouds are not virialized.

\subsection{Estimate From Surface Densities}
\label{mwsigma}

We can also compute the Milky Way star formation rate using surface
densities, the rotation curve, and the velocity dispersion. The vast
majority of star formation in the Milky Way occurs in a ring from 3 to
11 kpc in galactocentric radius \citep{mckee97} within which the
molecular and atomic gas surface densities are roughly
\citep{wolfire03}
\begin{equation}
\Smol \approx 
\left\{
\begin{array}{lr}
6.3 \exp\left(
-\frac{(r_k-4.85)^2}{ 2 \cdot 2.25^2}
\right)
\;\msun\mbox{ pc}^{-2},
&
3 \le r_k < 6.97 \\
4.1 \exp\left(-\frac{r_k-6.97}{2.89}\right)
\;\msun\mbox{ pc}^{-2},
&
r_k \ge 6.97
\end{array}
\right.
\end{equation}
and
\begin{equation}
\Sigma_{\rm HI} \approx
\left\{
\begin{array}{lr}
(2.0 r_k - 0.8)
\quad\msun\mbox{ pc}^{-2},
&
r_k < 4 \\
7 \quad\msun\mbox{ pc}^{-2}
&
4 \le r_k < 8.5 \\
\left[-1.57 + 8.57(r_k/8.5)\right]
\quad\msun\mbox{ pc}^{-2},
&
r_k > 8.5
\end{array}
\right.
,
\end{equation}
where $r_k$ is the galactocentric radius in kpc, and we have
multiplied the \citet{wolfire03} values for the surface density of
hydrogen by $1.4$ to get the total surface density including both H
and He. From these surface
densities we can directly compute $\Sg$, $\fGMC$, and $\phirho$. The
galactic rotation speed is $\vrot \approx 220$ km s$^{-1}$, and is
flat over the ring \citep{binney98}, so
\begin{equation}
\Omega = \frac{0.22}{r_k}\mbox{ Myr}^{-1}.
\end{equation}
The temperature in the molecular gas is $\sim 10$ K
\citep{solomon87}, giving a sound speed $\cs\approx 0.2$ km
s$^{-1}$. We estimate $\calm$ as a function of radius from $\Sg$ and
$\Omega$ using (\ref{macheqn}).

\begin{figure}
\epsfig{file=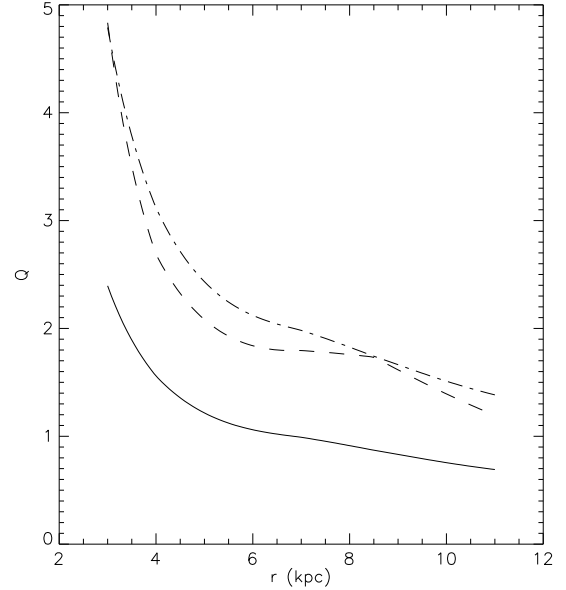}
\caption{\label{mwqrad}
Predicted value of $Q$ versus radius, estimatated using azimuthal
averages and scale heights (dot-dashed line), using azimuthal averages
and $\sg=6$ km s$^{-1}$ (dashed line), and corrected for spiral
structure (solid line).
}
\end{figure}

The final step is to estimate $Q$, which we do in two different
ways. First, we compute $Q$ from $\Sg$ assuming that the gas velocity
dispersion is $\sg=6$ km s$^{-1}$ independent of radius. This is
consistent with observations \citep{kennicutt89, heiles03}, although
the observations are quite uncertain because it is difficult to
determine the velocity dispersion as a function of radius within the
galaxy. Second, we compute $Q$ from the gas scale height, which can be
directly measured in the Milky Way. Equation (\ref{hgeqn1}) allows us
to compute $\rhog$ from $\Sg$ and $\hg$, and equation (\ref{rhogeqn2})
gives $Q$ in terms of $\rhog$ and $\Omega$. The scale heights of the
atomic and molecular gas within the star-forming ring are
\citep{wolfire03}
\begin{equation}
h_{\rm HI} \approx
\left\{
\begin{array}{lr}
65\mbox{ pc},
& \qquad
r_k < 8.5 \\
65 \,\exp\left[(r_k-8.5)/6.7\right]\mbox{ pc},
& \qquad
r_k \ge 8.5
\end{array}
\right.
\end{equation}
and
\begin{equation}
h_{\rm mol} \approx
\left\{
\begin{array}{lr}
33\mbox{ pc},
& \qquad
r_k < 8.5 \\
33 \,\exp\left[(r_k-8.5)/6.7\right]\mbox{ pc},
& \qquad
r_k \ge 8.5
\end{array}
\right.
.
\end{equation}
Note that we have converted the half-density heights given by
\citet{wolfire03} to scale heights by assuming an isothermal density
profile $\rho\propto \mbox{sech}^2[z/(2\hg)]$. We determine a
$Q$ by computing the midplane density of atomic and molecular gas, and
then solving (\ref{rhogeqn2}) for $Q$ using the surface density-weighted
average of the two midplane densities. The result agrees to within
$20\%$ with the value $Q$ as a function of radius we derive using the
first method. We plot the azimuthally-averaged $Q$ versus radius for
the Milky Way in Figure \ref{mwqrad}. However, most Milky Way star
formation occurs in the spiral arms. \citet{balbus88} shows that the
local $Q$ value in a spiral arm is related to the azimuthally averaged
$Q$ by
\begin{equation}
Q_{\rm arm} \approx Q_{\rm avg} 
\left(\frac{\Sigma_{\rm arm}}{\Sigma_{\rm avg}}\right)^{-1/2}.
\end{equation}
The Milky Way's spiral arms are overdense by factors of $\sim 4$
\citep{nakanishi03}, so we reduce our estimated value of $Q$ by a
factor of 2 to account for this effect. We also show the corrected $Q$ 
in Figure \ref{mwqrad}.

Integrating over the star-forming ring, we find a predicted star
formation rate
\begin{eqnarray}
\label{mwsfrpredrad}
\dot{M}_{\rm *-pred} & \approx &
\int_{3\;\mbox{\scriptsize kpc}}^{11\;\mbox{\scriptsize kpc}}
9.5 \;
\fGMC \,
\phipbarsix^{0.34}
\Qonefive^{-1.32} \,
\Omzero^{1.32} \,
\Sgtwo^{0.68} \nonumber \\
& & \qquad
2\pi R \, dR 
\quad\msun\mbox{ yr}^{-1}\mbox{ kpc}^{-2}
\\
& \approx & 4.5\quad\msun\mbox{ yr}^{-1}.
\end{eqnarray}
This agrees with the observed star formation rate of 3 $\msun$
yr$^{-1}$ in the Milky Way \citep{mckee97} and with our estimate based 
on observed GMC properties to better than a factor of 2. If we omit
the correction for spiral arms, we get a star formation rate of $2.1$
$\msun$ yr$^{-1}$, still in good agreement, so the spiral arm
correction is not critical.

\begin{figure}
\epsfig{file=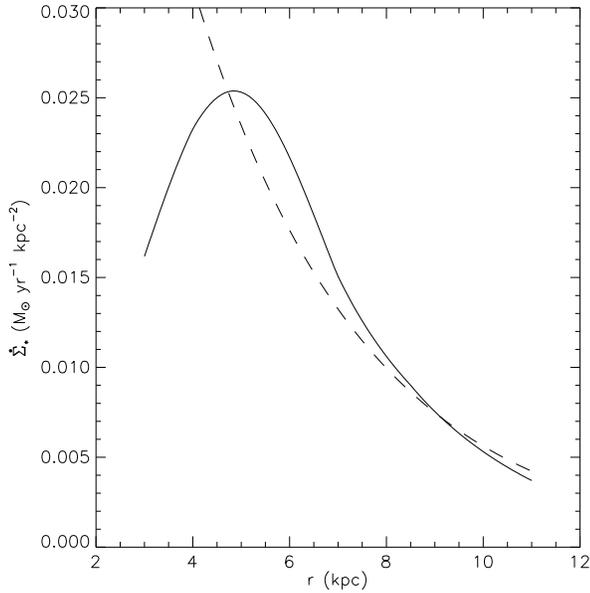}
\caption{\label{mwsfrrad}
Predicted variation in the star formation rate per unit area, $\Ssfr$, 
with galactocentric radius $r$. The solid line is our model, and the
dashed line is the model of \citet{mckee97}, scaled to have the same
integrated star formation rate that we predict.
}
\end{figure}

Note that (\ref{mwsfrpredrad}) gives a prediction not just for the
total star formation rate in the galaxy, but also for the radial
distribution of star formation. We show this in Figure
\ref{mwsfrrad}. For comparison, we also show the model of
\citet{mckee97} (scaled to have the same integrated star formation
rate as ours), which is generally consistent with observational data 
on the radial distribution of star formation outside 4 kpc. Our model
is similar to the McKee \& Williams model in this range, but differs
substantially inside 4 kpc because McKee \& Williams use a simple
exponential distribution with a cutoff for the radial variation of the
molecular gas surface density, while we use a more accurate
distribution that better reflects the decline in the molecular gas
surface density in the inner galaxy. We find that the
characteristic radius of star formation in the Milky Way, defined as
the radius within which half the star formation occurs, is $R_{\rm
char} \approx 7.1$ kpc. Taking the outer radius of the star forming
disk to be 11 kpc, this gives $R_{\rm char} = 1.3 R_{1/2}$. This
suggests that the common observational practice of measuring
quantities such as angular velocities at half the outer radius of star
formation \citep{kennicutt98b} should be reasonably accurate.

\section{Comparison to Galactic-Average Star Formation Rates}
\label{galaxycomp}

\subsection{Statistical Comparison}
\label{statcomp}

For a second test we compare our prediction against a sample of
95 galaxies, taken from the normal galaxies and
starbursts compiled by \citet{kennicutt98a} plus starbursts from
\citet{downes98}. For the Kennicutt galaxies,
we use the measured values of $\Smol$, $\Sg$, and $\tdyn$ as
reported in Tables 1 and 2 of \citet{kennicutt98a} to compute a
theoretical star formation rate from (\ref{sfr1}). We follow
Kennicutt in taking $\Omega=4\pi/\tdyn$ to be the typical value of
$\Omega$ in the star-forming region, and we exclude galaxies for
which there is no measured value of $\tdyn$. For starbursts where
there is no measured value of $\Sg$ (only $\Smol$) we assume $\fGMC
= 1$.

For the \citet{downes98} sources, we use a compilation of supporting
information from \citet{thompson05}. As with the Kennicutt starbursts, 
we take $\fGMC=1$ for all these points. We derive $\Sg$ and $\tdyn$
from the gas mass, half-power radii, and rotation curves from Tables
4, 5, and 9 of \citet{downes98}, and we derive star formation rates
from the FIR luminosities taken from the texts of \citet{downes98}
(IRAS00057+4021, IRAS02483+4302, VII Zw 31), \citet{genzel01}
(IRAS23365+3604, IRAS17208-0014), \citet{heckman00} (IRAS10565+2448),
and \citet{soifer00} (Mrk 231). We compute the star formation rates
from the FIR luminosities using the conversion factor of
\citet{kennicutt98b}. The data set includes multiple points for Arp 193, Mrk
273, and Arp 220 because \citet{downes98} break the sources up into a
more diffuse component and one or two ``extreme'' starburst
nuclei. For these objects we include both the diffuse component and
the nucleus or  nuclei. Data for the surface densities, dynamical
times, and luminosities for the diffuse components come from Tables 4,
5, and 9 and the text of \citet{downes98}, while data for the nuclei
come from Table 12.

To compare to this sample, we compute
\begin{equation}
\chi^2 \equiv \frac{1}{N_{\rm data}-N_{\rm fit}}
\sum [\log (\dot{\Sigma}_{\rm *-data}) -
\log (\dot{\Sigma}_{\rm *-theory})]^2
\end{equation}
for our model, and, as a normalization, for the
\citet{kennicutt98a} empirically determined best fit. The number of fit
parameters $N_{\rm fit}$ is unity for the Kennicutt best fit and zero
for our model.  We find $\chi^2 = 0.40$ for the best-fit of
\citet{kennicutt98a}, while our theoretical model gives
$\chi^2=0.55$. Note that these are not traditional $\chi^2$ 
goodness-of-fit statistics, since we are using the logarithm of the
data, we have no error bars for the measurements, and the dominant errors
(arising from extinction, an imperfectly known IMF, and
similar astrophysical uncertainties -- see Kennicutt 1998b) are
systematic and therefore highly non-Gaussian. Instead, the meaning
of this statistic is that $10^{\chi}$ is the RMS factor by which the
model errs in estimating the star formation rate. Thus, our results
corresponding to RMS errors of a factor of $5.6$ for our model and a
factor of $4.3$ for the Kennicutt fit. Given the factor of several
systematic uncertainties in the measured star formation rates, these
values are essentially identical.

\subsection{The Kennicutt-Schmidt Law}
\label{kscomp}

The Kennicutt-Schmidt Law correlates the star formation rate with
either $\Sg\Omega$ or $\Sg$, while our theory makes a prediction based 
on $\Sg$, $\Omega$, and $\fGMC$. From an intuitive physical
standpoint, one would be surprised if the star formation rate did not
depend on all three of our parameters to at least some degree.
Thus, the two forms of the
Kennicutt-Schmidt Law represent two ways of projecting a
four-dimensional space (consisting of $\Sg$, $\Omega$, $\fGMC$, and
$\Ssfr$) onto two dimensions. To compare our theory directly to these
laws, as opposed to the underlying data as we did in
\S~\ref{statcomp}, we must make some additional approximations. We
stress that we make these approximations only for the purposes of the
projection, and that the right way to test our theory is to use the
measured values of $\Sg$, $\Omega$, and $\fGMC$, as we did in
\S~\ref{statcomp}. We make them because they allow us to
show, in a relatively intuitive manner, why projecting the
four-dimensional data down to two-dimensions still allows such a good
fit to the observations.

Since neither version of the Kennicutt-Schmidt Law involves $\fGMC$,
we must estimate it in terms of $\Sg$ or $\Omega$.
\citet{wong02} and \citet{rosolowsky05b} find that the ratio
of molecular to atomic gas follows the approximate relation
\begin{equation}
\label{fGMCvsp}
\frac{\Smol}{\Sigma_{\rm HI}} \approx
\left(\frac{P_{\rm mp}/k_B}{2.5\times 10^4\mbox{ cm}^{-3}\mbox{
K}}\right)^{1.0},
\end{equation}
with about half a dex of scatter. Since $P_{\rm mp}$ is just a
function of $\Sg$ in our model, equation (\ref{fGMCvsp}) gives
\begin{equation}
\label{fGMCvsSg}
\fGMC \approx \left(1+0.025\,\Sgtwo^{-2}\right)^{-1},
\end{equation}
for our fiducial $\phip=3$. Note that most of the dynamic range of
the Kennicutt-Schmidt Law lies above $\Sg=100$ $\msun$ pc$^{-2}$, for
which $\fGMC \gtsim 0.98$, where have made the approximation that
most molecular gas is in GMCs. Thus, $\fGMC$ is almost constant over most
of the range of the the Kennicutt-Schmidt Law, which is part of the
reason that a projection of the data that neglects $\fGMC$ makes
little difference.

\begin{figure}
\epsfig{file=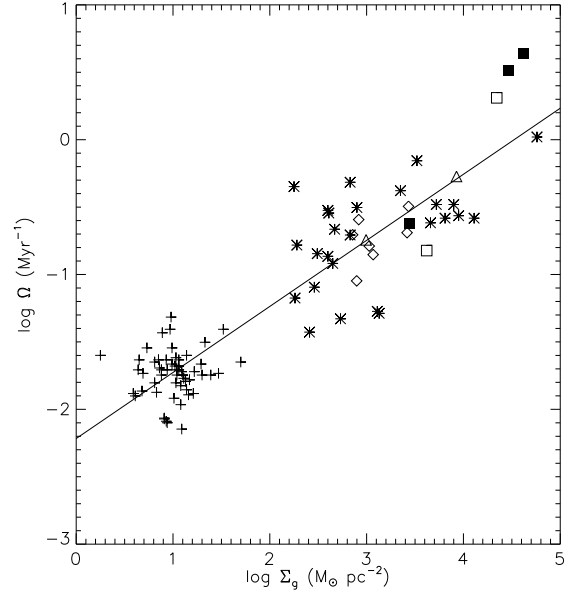}
\caption{\label{sigmaomega}
$\Sg$ versus $\Omega$ for observed galaxies. The data
points are: normal disks (crosses), circumnuclear
starbursts (asterisks), ULIRGs (diamonds), Arp 193 (triangles),
Markarian 273 (empty squares), and Arp 220 (filled squares). For a
description of how we derived these data points, see
\S~\ref{statcomp}. The line shows a linear fit to the data.
}
\end{figure}

With $\fGMC$ approximated in terms of $\Sg$, the remaining step is to
project from the three-dimensional space of $\Ssfr$, $\Sg$, and
$\Omega$ onto a two-dimensional space of $\Ssfr$ and just $\Sg$ or
$\Sg\Omega$. To do this, we make use of the fact that $\Sg$ and
$\Omega$ for galaxies appear to be correlated, as shown in Figure
\ref{sigmaomega}. The correlation is fit reasonably well by the rule 
\begin{equation}
\label{sigmaomegacorr}
\Omzero = 0.058\, \Sgtwo^{0.49},
\end{equation}
as the Figure shows. We can use this rule to estimate $\Sg$ and
$\Omega$ independently from any combination of $\Sg$ and $\Omega$,
allowing us to project our theory into the same lower-dimensional
space as the Kennicutt-Schmidt Law. This correlation is the other half 
of the reason that projecting the data into two dimensions works
well: $\Sg$ and $\Omega$ are not really independent, at least in the
available data set. Because they are correlated, projecting the data
onto any appropriately chosen combination of them will work, which is
why the $\Sg^{1.4}$ and $\Sg \Omega$ forms of the Kennicutt-Schmidt
Law work equally well, as does our prediction, which is approximately
$\Ssfr\propto \Sg^{0.68} \Omega^{1.32}$. Even though the parameter
space is four-dimensional, most of the data points lie near a line
within it, which makes distinguishing different models quite
difficult. We discuss how to break this degeneracy in
\S~\ref{futuretests}. Also note, however, that while
(\ref{sigmaomegacorr}) holds between galaxies, it is unknown if it
holds within galaxies. For this reason, the projection we derive to
compare to the Kennicutt-Schmidt Law may apply only to averages over
many galaxies, not within individual galaxies.

\begin{figure}
\epsfig{file=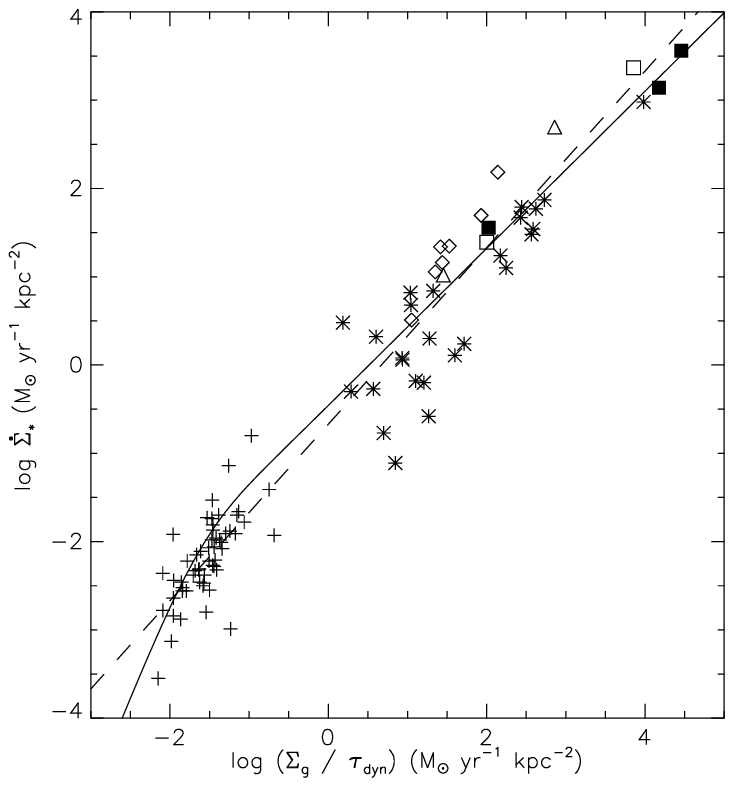}
\caption{\label{schmidt1}
Predicted star formation rate versus $\Sg / \tdyn$ (solid line). 
We also plot the \citet{kennicutt98b} best fit
(dashed line), and observed points for normal galaxies from
\citet{kennicutt98a} (crosses), circumnuclear starbursts from
\citet{kennicutt98a}
(asterisks), ULIRGs (diamonds), Arp 193 (triangles), Markarian 273
(empty squares), Arp 220 (filled squares). For a description of how
we derived these data points, see \S~\ref{statcomp}.
}
\end{figure}

\begin{figure}
\epsfig{file=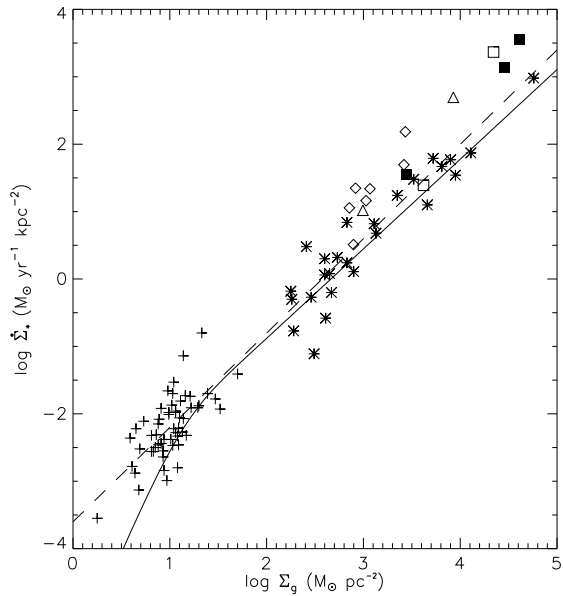}
\caption{\label{schmidt2}
Predicted star formation rate versus $\Sg$ (solid line). We
also plot the \citet{kennicutt98b} best fit (dashed line). The data
points are observed galaxies: normal disks (crosses), circumnuclear
starbursts (asterisks), ULIRGs (diamonds), Arp 193 (triangles),
Markarian 273 (empty squares), and Arp 220 (filled squares). For a
description of how we derived these data points, see \S~\ref{statcomp}.
}
\end{figure}

Using equations (\ref{fGMCvsSg}) and (\ref{sigmaomegacorr}) in
equation (\ref{sfr1}), our theoretical prediction for the star
formation rate in terms of $\Omega \Sg$ is
\begin{equation}
\Ssfr\approx 3.2 \,
\phipbarsix^{0.34} \,
\Qonefive^{-1.32} \,
\fGMC \,
\left(\Omzero \Sgtwo\right)^{0.89}
\;\msun\mbox{ yr}^{-1}\mbox{ kpc}^{-2},
\end{equation}
where
\begin{equation}
\fGMC \approx 
\left[1+5.5\times 10^{-3}\,
\left(\Omzero\Sgtwo\right)^{-1.34}
\right]^{-1}
\end{equation}
and $\phipbar\approx 10-8\fGMC$. The observed Kennicutt-Schmidt Law
with this choice of dependent variable is \citep{kennicutt98a}
\begin{equation}
\Ssfr=0.017 \, \Omega \Sg.
\end{equation}
We plot this and our theoretical prediction in Figure
\ref{schmidt1}. As the plot shows, our theoretical prediction, when we 
take into account the way that $\fGMC$, $\Sg$, and $\Omega$ are
related, essentially reproduces the first form Kennicutt-Schmidt Law. 
If we instead choose $\Sg$ to be our independent variable, following
the second form of the Kennicutt-Schmidt Law, our theoretical
prediction is
\begin{equation}
\Ssfr = 0.19 \,
\phipbarsix^{0.34} \,
\Qonefive^{-1.32} \,
\fGMC \Sgtwo^{1.33}.
\end{equation}
where $\fGMC$ is given by equation (\ref{fGMCvsSg}) and $\phipbar$ is
approximated in terms of $\fGMC$ as for the previous case.
The observed law is is \citep{kennicutt98a}
\begin{eqnarray}
\Ssfr & = & (2.5\pm 0.7) \times 10^{-4} \left(\frac{\Sg}{1\;\msun\mbox{
pc}^{-2}}\right)^{1.4\pm 0.15}
\nonumber \\
& & \qquad \msun\mbox{ yr}^{-1}\mbox{ kpc}^{-2} \\
& \approx & 0.16\, \Sgtwo^{1.4} 
\quad\msun\mbox{ yr}^{-1}\mbox{ kpc}^{-2}
\end{eqnarray}
We plot this and our theoretical prediction in Figure \ref{schmidt2},
and find that, again, our fit is reasonably
good. The only exception is at values of $\Sg \ltsim 10$ $\msun$
pc$^{-2}$. The error there arises from the fact that almost all the
galaxies with $\Sg$ so small lie well above the $\Omega$ versus $\Sg$
correlation we have used to project our theory (as shown in Figure
\ref{sigmaomega}), so the values of $\Omega$ we are using are
systematically smaller than those of the real galaxies in that region
of parameter space. Since our star formation rate depends on
$\Omega^{1.32}$, this underestimation of $\Omega$ causes the theory to
underpredict the star formation rate. If one uses the measured values
of $\Omega$ rather than the linear fit, the error at small $\Sg$ is no 
larger than it is elsewhere.

\section{Future Observational Tests}
\label{futuretests}

Our theory makes three observational predictions that should be
directly testable in the next few years. First, we can test our theory 
on nearby galaxies where molecular clouds are directly observable. In
\S~\ref{gmcpropsest} we compute the star formation rate in the Milky
Way by integrating over the observed distribution of Milky Way
GMCs. While we have some information about larger GMCs in nearby
galaxies, small GMCs make a non-negligible contribution to the star
formation rate there just as they do in the Milky Way. To reliably
compute the star formation rate in another galaxy, we must
therefore identify the lower mass cutoff below which molecular clouds
become non-virialized. This cutoff has not yet been observed in any
galaxy but the Milky Way, but such an observation is a straightforward 
extension of existing data sets to higher sensitivities and angular
resolutions. It should be within the
capabilities of the SMA, CARMA, or ALMA. Once one has determined the
full cloud mass distribution for another galaxy down to the non-virial 
cutoff, one can compute the star formation rate in another galaxy by
using (\ref{sfrdefn}) to compute the star formation rate in each
cloud, just as we have done for the Milky Way.
Since this type of test depends only on our calculation of
$\sfrff$, and not on any of our calculations of GMC properties in
external galaxies, this method allows our theory of $\sfrff$ to be
tested independently of the rest of our model.

Observations that resolve the star formation rate in a galaxy in
annular rings, but cannot resolve individual GMCs, provide a second 
possible test of our theory. With sufficiently good data, one
could use equation (\ref{sfrfit}) to predict the star formation rate
versus radius within a galaxy, just as we have done for the Milky
Way in \S~\ref{mwsigma}. This could then be compared to resolved
observations of star
formation versus radius, similar to those of \citet{wong02}. The
primary observational challenge in this comparison is that, to compare 
to a single galaxy where one cannot assume that parameters such as $Q$ 
have their average values, one must measure all the information that
we measured for the Milky Way. In particular, one must know $\Sg$,
$\fGMC$, $\Omega$, and $Q$ as a function of radius. The first three
are relatively straightforward, but measuring $Q$ requires that one be 
able to measure either the velocity dispersion or the gas scale
height. Neither quantity is easy to
determine observationally, but without it the theoretical predictions
will be uncertain by a factor of several. We suggest two possible ways
to make this measurement. First, one could perform resolved
observations of a starburst galaxy, where $\sg$ is large enough to be
comparable to the galactic rotation velocity and is therefore easier
to measure. Second, one could measure $\sg$ in a normal disk that is
face-on, and then use the \citet{tully77} relation to
obtain a rotation velocity. Since there is some scatter in the
Tully-Fisher relation, this procedure would likely need to be
performed over several galaxies to minimize the errors arising from
the uncertainty in the rotation curve.

A third possible test involves using a sample of galaxies similar to
but larger than that in \citet{kennicutt98a}. We found in
\S~\ref{kscomp} that, because $\Sg$ and $\Omega$ are themselves
correlated, $\Ssfr$ will correlate equally well with an infinite
number of combinations of $\Sg$ and $\Omega$. Our
theory predicts that the true scaling should be $\Ssfr\propto \fGMC
\Sg^{0.68} \Omega^{1.32}$, but the current data cannot distinguish
this combination of $\Sg$ and $\Omega$ from any other. However, there
is no reason that a future, larger sample of galaxies could not. In
order to break the degeneracy between combinations of $\Sg$ and
$\Omega$, a future sample must contain a large number of galaxies, or
annuli within galaxies, with fixed $\Sg$ and varying $\Omega$, or
fixed $\Omega$ and varying $\Sg$. With such a sample, one could
compute predicted star formation rates using (\ref{sfrfit}) and repeat
our analysis in \S~\ref{statcomp} and determine whether $\Sg^{0.68}
\Omega^{1.32}$ is a better fit. However, there is likely to be
considerable scatter arising from the stochastic nature of the cloud
and star formation process. This test will therefore require a
considerably larger sample of galaxies than are currently available.

Finally, note that one cannot easily test our theory by looking at
individual GMCs. Simulations of turbulence-regulated star formation
show significant fluctuations in the star formation rate versus time,
and we expect that real GMCs will also have large fluctuations. Thus,
our theory is valid only as an average over an ensemble of
GMCs. Furthermore, observations of a single GMC run into a problem
with GMC ages. Tracers of star formation such as FIR and radio
continuum luminosities measure the mass of stars formed over some
period in the past. The amount of time depends on the tracer, but even 
tracers sensitive only to the very youngest stellar populations
integrate the star formation over several Myr. We do not know the GMC
lifetime well, and even in our model of virialized GMCs we cannot rule
out the possibility that it is only $\sim 10$ Myr, a few GMC crossing
times. Thus, one cannot be confident when observing a single GMC that
it has been forming stars for long enough to have reached a
steady-state luminosity in the tracer that one is using. This makes
observations difficult to interpret, because one cannot break the
degeneracy between the star formation rate and the age of the cloud.

\section{Discussion}
\label{discussion}

\subsection{Estimate of Uncertainties}
\label{uncertainties}

We begin to estimate our uncertainties by considering how much our
estimates of the star formation rate could be off by considering a
``worst-case scenario'' for our unknown parameters, $\avir$, $\phip$,
$\phipbar$, $Q$, and $\ec$. Our fiducial value for $\avir$ is 1.3,
and a plausible range based on the observations is $1-2$.
As discussed in Appendices \ref{phipapp} and \ref{phipbarapp}, the
plausible ranges in $\phip$ and $\phipbar$ are $\phip=1-6$ and
$\phipbar=2-10$. We have also
used a fiducial value of $Q=1.5$. Simulations of purely gaseous
magnetized disks show that collapse in a disk can set in at $Q$ in the
range $0.9-1.6$ \citep{kim01, kim02, kim03}. Analytic work shows that
stars allow smaller values of $Q$ to be stable
\citep{jog84,rafikov01}. Observationally, most galaxies fall in the
range $Q=0.75-3$ \citep{martin01, wong02}, with outliers going as far
as $Q=0.5$ to $Q=6$. We adopt $Q=0.75-3$ as our plausible range of
variation for most galaxies. Finally, we have taken $\ec=0.5$, but the
plausible range for the mass fraction ejected by feedback is
$\ec=0.25-0.75$ \citep{matzner00}.

If we consider all of these parameters simultaneously assuming their
extreme values, for a given galaxy we can reduce our predicted star
formation rate by as much as a factor of $10$, and increase it by as
much as a factor of $6$, relative to our fiducial case given by the
parameters in Table \ref{paramlist}. A more realistic estimate of the
error is probably a factor of $\sim 3$, because there is no reason our
errors should add up systematically in this fashion. Indeed, the
maximum errors are possible only for combinations of parameters that
can be ruled out on observational grounds other than the
Kennicutt-Schmidt Law. For example, a
reduction of the star formation rate by a factor of $10$ is possible
only if $\phipbar=2$, $\phip=6$, and $\avir=2$, giving
$\phirho=1.3$. This corresponds to a galaxy where molecular clouds are
only overdense relative to the mean in the ISM by $30\%$. No known
galaxy, including ones where the ISM is entirely molecular, has
clouds with such a small overdensity compared to the rest of the
ISM. Indeed, such a galaxy would effectively have no clouds at all,
just a continuous intercloud medium. Similarly, an increase
in the star formation rate by a factor of $6$ occurs for
$\phipbar=10$, $\phip=1$, and $\avir=1$. Plugging in Milky Way values
of $\Omega$ and $\Sg$ with these parameters gives $\calm\approx 5$,
much smaller than the observed velocity dispersion in GMCs in the
Milky Way or in any other galaxy.

We can also identify a number of uncertainties not associated with any 
specific parameters, but instead with conceptual assumptions that we
have made. First, observations have confirmed that, in at least some
clouds, the star formation rate is lower in the outer than the inner
parts \citep{li97, johnstone04}, perhaps due to increased ionization
there \citep{mckee89}. The periodic box simulations we have used to
calibrate $\sfrff$ do not include any effects arising from the finite
size of real GMCs, and this may produce an error.
A second effect is that we have assumed that all the gas in starbursts
is in bound structures capable of forming stars. However, observed
galactic nuclei and starbursts that are molecular throughout
consist of a collection of clouds with a molecular intercloud medium
\citep{solomon97, rosolowsky05}. Our assumption that all the gas is in
bound structures may therefore cause us to systematically overestimate
the star formation rate. However, \citet{rosolowsky05} find that in
M64 the clouds account for $\sim 75\%$ of the mass, and simulations
such as those VBK03 of show that $\gtsim 50\%$ of the mass does
collapse in unstable environments, so the error is probably small. 
Third we have neglected magnetic fields. We argue in
\S~\ref{magdiscussion} that star-forming clouds are likely magnetically
supercritical, and thus cannot be held up against collapse by magnetic 
fields, and we present preliminary evidence in \S~\ref{simcomparison}
that a magnetic field in a fairly supercritical cloud does not
substantially inhibit star formation. However, it is possible that a
magnetic field stronger than the one used in \citet{li04}, yet still
not strong enough to make the cloud subcritical, could inhibit the
formation of cores by preventing gas from flowing across field lines
to accrete onto them. We find this unlikely, however, since in a
supercritical cloud the Alfv\'en Mach number is likely to be unity or
greater. Fourth, we have ignored the possible effects of star
formation in objects like the clumps observed by \citet{plume97} that
do not lie on the linewidth-size relation, and that numerical
simulations thus far lack the resolution to model. Since these objects
are over-pressured and  over-dense compared to typical galactic
star-forming clouds, they have shorter free-fall times and form stars
faster. By neglecting them, we probably systematically underestimate
the star formation rate. The extent of the underestimate is somewhat
uncertain, since simulations to date have not modeled this effect,
and we do not know how much exactly mass is in these clumps in the
galaxy.

\subsection{Application to Simulations}

Our theory of turbulence-regulated star formation is readily
applicable to simulations on cosmological or galactic scales that do
not have enough resolution to model molecular cloud formation or star
formation directly. This is particularly true because, while we can
integrate over an entire galactic disk to compute average star
formation rates, we also predict the star formation rate in terms of
local properties of the gas.

In a simulation, one usually wants to implement star formation as a
sub-grid model. This requires a recipe for determining at what rate
the mass in a given cell or particle is transformed into
stars. Equation (\ref{sfrdefn})
gives the star formation rate in terms of the local free-fall
time $\tff$, molecular mass $M_{\rm mol}$, and the star formation rate 
per free-fall time $\sfrff$, which is a function of $\avir$ and
$\calm$, the local virial parameter and Mach number. Since in a
simulation the density of every cell is generally known, it is simple
to compute $\tff$. Since the gas mass but not the molecular mass of
every cell is known, one must determine $\fGMC$ to find $M_{\rm
mol}$. To do this, one may either assume that sufficiently dense cells
are entirely molecular, or more directly use the observed relation
between pressure and $\fGMC$, equation (\ref{fGMCvsp}). (One should be
wary of applying this rule to galaxies with metallicities too
different from that of the Milky Way, though, since the correlation
almost certainly has some metallicity dependence.)

Finally, to compute $\sfrff$, one needs to know the virial parameter
and Mach number within a cell. The easiest way to estimate $\calm$
is to compute the velocity dispersion over a small region around
the cell, and extrapolate down to the size of the cell using the
linewidth-size relation $\sigma\propto l^{0.5}$. This plus the
temperature of the cell yields $\calm$ within the cell. While this
procedure is somewhat uncertain because it requires extrapolation to
scales below the grid size, $\calm$ has only a weak effect on
$\sfrff$. One can compute $\avir$ by using $\calm$ to estimate the
kinetic energy in the cell, and comparing to the estimated
gravitational self-energy of the cell. This process for estimating
$\avir$ is similar to the process of estimating whether a region is
bound used in the sink particle creation procedures outlined
by \citet{bate95} for Lagrangian codes and \citet{krumholz04} for
Eulerian codes. From $\avir$ and $\calm$, one can compute $\sfrff$,
and from that $\dot{M}_*$.

This procedure provides a simple estimate for the rate at which a cell
turns its mass into stars that is based on a physical model rather
than an arbitrary density cutoff and efficiency for star formation,
which are commonly used in simulations now. One caveat on our
approach, however, is that it does not apply to primordial star
formation, where the primary limit on star formation is the ability of 
the gas to cool, rather than turbulent support.

\subsection{Magnetic Fields}
\label{magdiscussion}

Our theory of star formation regulated by supersonic turbulence is
only valid if star formation occurs primarily in regions
that are magnetically supercritical. If molecular clouds are 
magnetically subcritical, then magnetic fields can prevent collapse,
and the time required for the flow to ``replace'' collapsing cores is
the ambipolar diffusion time rather than the free-fall time. This
effect could inhibit flow in unbound regions of GMCs even if the
clouds overall are supercritical. However, our comparison with the work of
\citet{li04} gives preliminary evidence that this effect is small.

Both observations and general theoretical considerations support the
idea that molecular clouds are
supercritical. Theoretically, \citet{mckee89} and \citet{mckee93}
point out that GMCs cannot be bound, turbulent, and magnetically
subcritical. Turbulence and magnetic fields together can support a
larger cloud mass than magnetic fields or turbulence alone. If the
turbulent energy is comparable to the magnetic energy, as both
observations and general expectations of equipartition suggest, then
the critical mass arising from both sources must be $M_{\rm crit}
\approx 2\mphi$. If the cloud is magnetically subcritical, then $M
\ltsim \mphi$, so $M\ltsim M_{\rm crit}/2$. However, for a cloud to be
bound it must be near its critical mass. A cloud that is only half its 
critical mass must be unbound, and would certainly not be
centrally concentrated. Since observations indicate that molecular
clouds are both bound and centrally concentrated (see
\S~\ref{bounddiscussion}), it follows that they must be magnetically
supercritical, with $M\approx 2 \mphi$.

Zeeman splitting observations of magnetic field strengths in Milky Way
GMCs support this view. \citet{crutcher99} and \citet{bourke01} find
that $M \approx 2\mphi$. \citet{galli99} and \citet{allen00} point out
that this conclusion depends on the assumed cloud geometry along the
line of sight, and that Crutcher's data are consistent with $M \approx
\mphi$ if clouds are highly flattened. However, this model is feasible
only if true magnetic field values in regions with no detectable
Zeeman splitting are near their $3\sigma$ upper limits
\citep{bourke01}. Furthermore, if GMCs in general are highly flattened,
we ought to observe at least some of them edge-on, allowing us to see
their sheet-like structure. No such sheet-like clouds have been
observed. A final problem with the sheet-like cloud picture is that
a highly flattened geometry is not consistent with the observation
that clouds have turbulent energies comparable to their gravitational
potential energies. In such a cloud, the turbulence would be strong
enough to bring gas out of the cloud plane and create a more
three-dimensional geometry.

Another line of observational evidence that clouds are
magnetically supercritical
comes from statistical indicators. \citet{padoan04b} argue based on
simulations that the magnetic fields that are at or above
equipartition with the kinetic energy yield measurably different
distributions of column density than fields that are below
equipartition. They argue that the observations are closer to the
sub-equipartition simulations. While there is some uncertainty in
interpreting simulations of periodic boxes in the context of real,
finite-sized molecular clouds, these simulations to provide a strong
argument for magnetic supercriticality.

One final problem for magnetically-mediated star formation theories
is that the time required for ambipolar diffusion to change a
subcritical region into one that is supercritical may be considerably
shorter in turbulent media than in static media
\citep{heitsch04}. Consequently, the long ambipolar diffusion time
invoked to explain the low SFR may not apply to GMCs, which
observations indicate are strongly turbulent. \citet{zli04} and
\citet{nakamura05} perform simulations showing that in a
two-dimensional geometry, turbulence does not enhance ambipolar
diffusion enough to make the star formation rate too high, but
two-dimensional and three-dimensional turbulence are very different,
so it is unclear that their results in this regard are applicable to
real clouds. Whether magnetic regulation with ambipolar diffusion is
even capable of producing the correct star formation time scale in a
turbulent medium remains an open question.

\subsection{Why Is $Q \approx 1$?}

An additional important assumption in our theory is that $Q\approx
1$. While this is well-justified observationally \citep{quirk72,
kennicutt89, martin01}, previous work has
also provided a theoretical explanation, which is part of any complete
theory of star formation. Theoretically one expects feedback effects
to prevent $Q$ from straying too far from unity. In ordinary disk
galaxies like the Milky Way, supernovae are the likely feedback
mechanism \citep{silk97}. If $Q$ is too low then the star formation 
rate will increase (equation \ref{sfr1}) and the supernova rate will
increase as well. This will raise the temperature and the velocity
dispersion in the ISM, increasing $Q$ and reducing the star formation
rate. If $Q$ becomes too large compared to unity, then gravitational
instability shuts off and molecular clouds cease to form. This is
observed in the outer parts of disk galaxies
\citep{kennicutt89}. However, if there is sufficient gas present, then
without continual supernova stirring the gas velocity dispersion
will decrease. This will reduce $Q$, causing the star formation rate
to rise again.

In starbursts, the feedback mechanism probably changes over from
supernovae to radiation pressure \citep{thompson05}, but the effect is 
similar. Low $Q$ values raise the star formation rate, which increases 
the luminosity of the stellar population and thereby increases the
radiation pressure. This puffs up the disk and restores $Q\approx
1$. If $Q$ is much larger than unity, the star formation rate will
fall and the disk will lose radiation pressure support and begin to
collapse, reducing $Q$. These mechanisms complete the picture of why
$Q\approx 1$.

\subsection{Are Molecular Clouds Bound?}
\label{bounddiscussion}

Our analysis also depends on molecular clouds being gravitationally
bound, virialized structures. If the true virial parameter of GMCs in
substantially different from unity, then $\sfrff$ and the overall star
formation rate will be much greater (for $\avir \ll 1$) or smaller
(for $\avir \gg 1$) than we have estimated. Furthermore, our analysis
based on the density probability distribution function assumes that
molecular clouds are gravitationally bound structures that live long
enough for their density distributions to reach statistical
equilibrium. If GMCs are largely unbound, or consist of gas that has
been compressed by shocks and that all collapses immediately
\citep{elmegreen00, hartmann01, clark04, clark05}, it is not clear
that the density PDF could reach its equilibrium form before the star
formation process was complete. We must therefore consider whether our
assumption of bound, virialized clouds is a sound one.

Observations indicate that GMC virial parameters are close to
unity. \citet{mckee03} analyze the CO surveys of \citet{solomon87} and
\citet{dame86}, and find that the mean virial parameters for the large
clouds in their samples, where most stars form, are 1.3 and 1.4. In
M33, \citet{rosolowsky03} obtain velocity dispersions, masses, and
radii for 36 GMCs. From their data, we find a mass-weighted mean
virial parameter of 1.6. In the nucleus of M64, \citet{rosolowsky05}
find that GMCs are overpressured with respect to their environments by
at least a factor of 2, indicating that they too likely have
$\avir\approx 1$. Thus, our adopted value of $\avir=1.3$ is in good
agreement with observations, both in the Milky Way and in the disks
and nuclei of galaxies similar to it.

That observed virial parameters are all close to unity in itself
strongly indicates that GMCs are gravitationally bound, not held
together temporarily by the ram pressure of turbulent flows in the
ISM. There is no reason that turbulent flows would create clouds with
$\avir\approx 1$. As an example, consider the molecular clumps inside
GMCs, most of which are created by turbulent flows and confined by
turbulent pressure rather than self-gravity. Most clumps have virial
parameters $\avir\gg 1$, and they have a power-law distribution of
$\avir$ values for $\avir \gtsim 1$ \citep{bertoldi92}. The same is
true of molecular clouds with masses $\ltsim 10^4$ $\msun$
\citep{heyer04}. While 
molecules will only form in dense regions of 
the ISM, and for this reason CO surveys are biased towards dense gas
with low virial parameters, for the UV field of our galaxy to be such
that we see only clouds that have a virial parameters of $1-2$
requires an unlikely coincidence. Even if this coincidence could work
in the Milky Way, it would not explain the observations in M33, where
the interstellar UV flux could be quite different, and in M64, where
the density of the gas prevents FUV photons from propagating through
the ISM at all.

Another strong argument that suggests GMCs are bound is that GMCs have 
a characteristic mass. In the Milky Way, there is a clear upper limit
on GMC masses of approximately $6\times 10^6$ $\msun$. This limit is
not consistent with statistically ``running out'' of clouds at high
masses. It is a real break in the power-law distribution that is
observed at lower masses \citep{mckee97}. The mass distributions of
GMCs in M33 \citet{engargiola03} and
M64 (E. Rosolowsky, private communication) also
exhibit characteristic scales. If GMCs are gravitationally bound, then
the Jeans mass provides a natural scale that agrees reasonably well
with the observations. If GMCs are not bound, however, they cannot
have been created by gravitational collapse and the Jeans mass is
therefore irrelevant. Turbulent flows without self-gravity are
scale-free. If GMCs are unbound they should not exhibit any
characteristic mass. This prediction of the unbound GMC model is
inconsistent with the observations. One cannot invoke observational
selection biases to explain this inconsistency, as is done to explain
the observed values of $\avir$. Rendering the Milky Way GMC mass
distribution consistent with a pure power law would require that the
Milky Way contain $\approx 100$ GMCs with masses larger than $6\times
10^6$ $\msun$ \citep{mckee97, mckee99}. There is no plausible way that 
such a large number of very massive clouds could have been missed.

\subsection{Feedback and Cloud Destruction}
\label{feedback}

Thus far we have omitted any discussion of the effects of massive star 
formation feedback. Obviously massive star formation gives rise to
HII regions that destroy molecular clouds by photoionization and
winds. \citet{matzner02} estimates that this effect limits galactic
GMCs to converting at most $\sim 5-10\%$ of their mass into stars over
their lifetimes. Our justification for neglecting this effect hinges
on the difference between the star formation \textit{efficiency},
which measures the fraction of gas in a particular GMC that is
transformed into stars, and the star formation \textit{rate}, which
measures the instantaneous rate at which gas is transformed into
stars. Feedback from massive stars ultimately controls the star
formation efficiency by disrupting a cloud before it can turn most of
its mass into stars. However, feedback does not change the
instantaneous star formation rate in the molecular gas except
indirectly, by driving turbulence in the molecular gas and therefore
changing the Mach number. Feedback only affects the star formation
rate by turning molecular gas into atomic or ionized gas, thereby
reducing the amount of molecular gas available to make stars.
A thorough understanding of
mechanisms like photoionization that regulate the amount of molecular
gas available to form stars would allow us to calculate $\fGMC$ from
first principles, rather than taking it from observations, and would
be an important piece of a complete theory of star formation. However, 
our results can stand independently of this, since $\fGMC$ is directly 
observable, and our theory therefore relies only on direct observables.

\subsection{Turbulent Decay}
\label{turbdecay}

The largest single omission from our theory of star formation is that
it does not address the critical question of what keeps GMCs in virial 
balance. Simulations of both hydrodynamic and magnetohydrodynamic
turbulence in periodic boxes indicate that turbulence decays on time
scales of a single crossing time of the object \citep{maclow98, stone98,
maclow99, padoan99}, The crossing time is $\tcr=2R/\sigma$, where $R$
is the object's radius and $\sigma$ is its velocity dispersion. The
crossing time and the free-fall time are related to the virial
parameter by
\begin{equation}
\label{ffvscr}
\avir = \frac{5\sigma^2 R}{G M} = \frac{40}{3\pi}
\left(\frac{\tff}{\tcr}\right)^2,
\end{equation}
where $M$ is the object's mass. In a cloud with our fiducial value of
$\avir=1.3$, $\tcr=1.8\tff$. If the turbulence decays substantially in
a single crossing time, this means that the object should enter
free-fall collapse within $\sim 2$ free-fall times. If that happened,
then $\avir$ would become much smaller than unity, and the majority of
the gas would rapidly turn into stars. That would yield a star
formation rate far higher than observations allow. Thus, GMCs must not
be collapsing in this manner. Rapid decay of turbulence is also
difficult to reconcile with several other observations (see McKee 1999
for a detailed discussion).

There are several possible explanations for the non-collapse of
GMCs. First is the possibility that turbulence may not decay as
quickly as the simulations indicate. \citet{cho03} argue that Alfv\'en 
waves in a turbulent magnetized medium cascade from large to small
scales and decay anisotropically, with modes along and perpendicular
to the magnetic field having different decay rates. Only one mode
decays as rapidly as  the simulations indicate. They argue that the
simulations performed to date lack the dynamic range to model this
effect correctly. Similarly, \citet{sugimoto04} perform simulations
showing that, in a filamentary cloud geometry, Alfv\'en waves of
different polarizations decay at different rates, with some modes
decaying twice as slowly as earlier simulations indicated. If these
results from somewhat idealized cases apply to real clouds, then GMCs
could live for several free-fall times, long enough to allow the
formation of massive stars that could disrupt them rather than
letting them collapse entirely into stars.

A second possibility is that turbulence in GMCs is driven by continual 
perturbations from outside that are strong enough to prevent
the decay of turbulence and keep GMCs virialized. \citet{kornreich00}
suggest that GMCs will be struck by supernova shock waves that
maintain cloud turbulence at intervals comparable to the free-fall
times of large GMCs. However, this source of driving is highly
stochastic, so it is unclear the shocks can truly keep most clouds
from collapsing. Furthermore, \citet{nakamura05b} perform numerical
studies  indicating that it may not be possible for external shocks to
drive turbulence in clouds without disrupting them entirely. 
\citet{koyama02} suggest that turbulence is driven by thermal
instability at the interface between atomic and molecular
gas. However, the characteristic size scale of the disturbances this
creates is only $\sim 0.1$ pc, so it is unclear that this turbulence
would be able to affect the interiors of GMCs. \citet{piontek04}
consider thermal plus magnetorotational instability in the atomic
phase of the ISM, and find that magnetic fields allow motions
generated at the warm-cold interface to drive turbulence far from the
interface. However, it is unknown if this mechanism would work in GMCs.
Furthermore, thermal instability offers no clear way to explain
turbulence in GMCs in galaxies like M64 where the ISM has no atomic
phase, and is  not known to be thermally bistable as is the atomic ISM
in the Milky Way.

A third possible solution to the problem of turbulent decay is driving 
by feedback from star formation. \citet{norman80} and \citet{mckee89}
argue based on analytic calculations that, for the observed star
formation rate, the rate at which protostellar outflows inject energy
into their parent clouds is sufficient to balance the rate at which
turbulence decays. \citet{quillen05} observe protostellar outflow
cavities in NGC 1333, and estimate that the rate of energy injection
from the observed cavities is sufficient to power the turbulence of
the cloud, in agreement with this model. \citet{matzner02} argues that
when massive stars are present, turbulent motions driven by
the overpressure in HII regions are the dominant source of energy
injection. Matzner estimates analytically that the energy injection
rate by HII regions is sufficient to balance the turbulent decay rate
even if the decay time is only a crossing time. However, the theory
depends on an efficiency of energy injection by HII regions that has
only been estimated analytically, and ideally should be set by
simulations.

Regardless of the true mechanism, the observations show that GMCs
cannot be collapsing completely and rapidly. The exact mechanism by
which the turbulence is maintained does not affect our analysis,
because, below the scale at which it is driven, all turbulence is the
same. That is why, for example, simulations find a universal density
probability distribution function independent of whether the
turbulence is driven or undriven, and regardless of the random
realization of the initial velocity field or driving field.
Observed GMCs both in the Milky Way and in other galaxies are
virialized, with turbulence balancing gravity, and we have shown here 
that virialized, turbulent clouds produce a star formation rate that is 
consistent with observations. The remaining significant piece of this
theory, which we leave for future work, is an explanation for how the
observed virial balance is maintained.

\section{Conclusion}
\label{conclusion}

In this work we have attempted to fill in a significant missing piece
of the overall picture of star formation: a quantitative theory that can
map the conditions in a star forming region into a star formation rate 
based on simple physical principles. Our basic picture is that
stars form in gravitationally bound, virialized molecular clouds. Only
$1-2\%$ of a cloud is transformed into stars in a single free-fall
time because in a turbulent virialized cloud, most of the gas is in
structures that have more kinetic energy than gravitational potential
energy. Only rare, overdense regions are gravitationally bound, and
the fraction of a cloud's mass in such regions is nearly a constant
$\sim 1\%$ over all virialized clouds. We have for the first time
computed the collapsing mass fraction directly in terms of the Mach
number and the virial parameter, the two basic dimensionless numbers
that describe a star-forming cloud, and shown that the fraction of gas
in collapsing structures is only a very weak function of the Mach
number for virialized clouds. The star formation rate is simply the
mass in sufficiently overdense structures divided by the cloud
free-fall time. Our model does not rely on an unknown efficiency
of star formation or an unknown critical density. The only inputs are
the physics of turbulence and the virial theorem.

This prescription correctly predicts the star formation rate when we
apply it to the observed giant molecular clouds in the Milky Way. We
also estimate the properties of star-forming clouds in other galaxies
as a function of the rotation speeds and surface densities of various
component in those galaxies. We use these estimated cloud properties
combined with our prediction for the star formation rate in a cloud to 
compute galactic-average star formation rates, and show that our
predictions agree with the observed star formation rate in a sample of 
galaxies ranging from normal disks like the Milky Way to starbursts
and ULIRGs. Thus, our theory provides a unified model capable of
explaining the star formation on scales from the individual clouds
within a galaxy to the entire star-forming disk of a starburst or
normal disk galaxy.

\acknowledgements The authors thank Leo Blitz, Norm Murray, Eliot
Quataert, Eric Rosolowsky, Jonathan Tan, and Todd Thompson for helpful
discussions. CFM acknowledges the support of NSF grant AST-0098365.

\begin{appendix}

\section{Estimating $\phip$}
\label{phipapp}

We estimate $\phip$ by considering 
cases ranging from normal disks to starbursts. In the solar neighborhood,
the total disk surface density is $\Stot \approx 56$
$\msun$ pc$^{-2}$ \citep{holmberg04}, and the gas surface density is
$\Sg \approx 12$ $\msun$ pc$^{-2}$ \citep{boulares90}, so $\fg\approx
0.21$. The total midplane pressure is $P\approx 3.9 \times 10^{-12}$
dyn cm$^{-2}$, but approximately $1.9 \times 10^{-12}$ dyn cm$^{-2}$
of this comes from magnetic fields and cosmic rays
\citep{boulares90}. Since these permeate the molecular clouds and the
non-molecular gas equally, they provide no confining pressure on
molecular clouds. The effective pressure on GMCs in the Milky Way,
therefore, is roughly $2 \times 10^{-12}$ dyn cm$^{-2}$. For the Milky
Way solar neighborhood values of $\Stot$ and $\Sg$, we find $\phimp =
0.50$. Thus, $\phip \approx 2.4$ in the solar neighborhood.

At the opposite extreme consider a starburst or
ULIRG. \citet{downes98} find that the gas fraction in high-surface
density starbursts is $\fg \approx 1/3$. We cannot directly observe
$\phimp$ in starbursts, but we can estimate it based on physical
considerations. The reason $\phimp < 1$ in the Milky Way is
that the gas scale height is small compared to the stellar scale
height. This occurs because the gas comprises a small fraction
of the total surface density of the disk, and because old stars have
had a long time to scatter off molecular clouds \citep{rafikov01}. In
a starburst, the gas fraction is 
considerably higher and there is no population of old stars that have
had a long time to be dynamically heated \citep{downes98}. We
therefore expect that stars and gas will have comparable scale
heights, which will produce $\phimp\approx 1$. This gives $\phip=3$ in 
starbursts.

Since $\phip$ seems roughly constant over a range of environments from 
the solar neighborhood to extreme starbursts, we adopt a constant
value of $\phip=3$ throughout our work. The plausible range of
variation of $\phip$ is from $\sim 1$, corresponding to a purely
gaseous disk, to $\phip\sim 6$, corresponding to a starburst
containing only $1/6$ gas, the rough lower limit in the
\citet{downes98} sample.

Note that, because GMCs occupy a relatively
small fraction of the ISM, one might treat them as a pressureless
component like stars rather than a pressure-contributing component
like atomic gas. This would reduce $\phip$. However, since within 
a GMC the molecular gas does contribute pressure, the product $\phip
\phipbar$ must remain unchanged. Thus, if one takes a smaller value
for $\phip$ one must use a correspondingly larger value for
$\phipbar$. Since our predicted star formation rate depends on $\phip
\phipbar$, there would be no net change to our predictions.

\section{Estimating $\phipbar$}
\label{phipbarapp}

In an environment where the ISM is predominantly atomic, such as the
Milky Way, interstellar UV photons dissociate
H$_2$ and CO that is not sufficiently shielded. Thus, molecular clouds
exist only as the inner parts of atomic-molecular complexes
\citep{elmegreen89, elmegreen94}. Since atomic and molecular hydrogen
cannot cool effectively, star formation only occurs in the parts of
the complexes 
where CO is present. For Milky Way interstellar UV fluxes, a layer of
gas where C is atomic must provide at least $\sim 0.7$ mag of
extinction to prevent dissociation of CO \citep{vandishoeck88}. With
such a shielding layer, the mean pressure in the molecular gas is
higher than in the combined atomic and molecular
complex. \citet{holliman95} estimates $\phipbar\approx 8$, which is
consistent with the observed ratios of GMC pressure to ISM pressure in
the Milky Way \citep{blitz93}. However, there is considerable
uncertainty in applying this estimate to other galaxies, because it
depends on the metallicity of the galactic ISM and the strength of the
interstellar UV field, both of which vary considerably from galaxy to
galaxy.

For galaxies where the ISM is purely molecular, clouds are not exposed
to any external UV flux. In this case, we assume that clouds can be
described very roughly as polytropic spheres. For a polytropic cloud
with $P\propto r^{-k_P}$,
\begin{equation}
\phipbar = \frac{3}{3-k_P}.
\end{equation}
For an isothermal sphere, $k_P = 2$ so $\phipbar=3$. For a cloud with
a density profile $\rho \propto r^{-1}$, $\phipbar=1$. We consider
these extreme limits, and take $\phipbar=2$ as a typical value. This
is consistent with observations of GMCs in purely molecular galaxies
\citep{rosolowsky05}.

We adopt a very rough formula to interpolate between the purely atomic 
and purely molecular cases:
\begin{equation}
\phipbar = 10 - 8\fGMC,
\end{equation}
where $\fGMC\equiv\Smol/\Sg$ is the molecular gas fraction. One could
also have chosen to use a step function approximation or simply taken
$\phipbar=6$ as a universal value covering the range from starbursts
to ordinary disks. We consider any value of $\phipbar$ from $2$ to
$10$ reasonable, although a value of $2$ is implausible for a galaxy
with a great deal of atomic gas, and a value of $10$ is implausible
for a galaxy that is entirely molecular.
\end{appendix}

\end{document}